\documentclass[iop]{emulateapj}
\usepackage{graphicx}
\usepackage{amssymb}
\usepackage{ctable}
\usepackage{natbib}

\def\Mpc{$h^{-1}$Mpc}
\def\mnras{MNRAS}
\def\ssr{SSRv}
\def\apj{ApJ}

\def\apjs{ApJS}
\def\apjl{ApJL}

\def\aap{A\&A}
\def\aj{AJ}

\shorttitle{Identifying members in the outer regions of clusters}
\shortauthors{Serra $\&$ Diaferio}

\begin{document}

\title{Identification of members in the central and outer regions of galaxy clusters}

\author{Ana Laura Serra}
\affil{
Istituto Nazionale di Astrofisica (INAF), Osservatorio Astronomico di Torino, \\
Strada Osservatorio 20, I-10025, Pino Torinese, Torino, Italy\\
Dipartimento di Fisica, Universit\`a di Torino, Via P. Giuria 1, I-10125,
 Torino, Italy \\
Istituto Nazionale di Fisica Nucleare (INFN), Sezione di Torino, Torino, Italy\\}
\email{serra@ph.unito.it}

\author{Antonaldo Diaferio}
\affil{
Dipartimento di Fisica, Universit\`a di Torino, Via P. Giuria 1, I-10125,
 Torino, Italy \\
Istituto Nazionale di Fisica Nucleare (INFN), Sezione di Torino, Torino, Italy\\}

\begin{abstract}
The caustic technique measures the mass of galaxy clusters in both their virial and infall regions
and, as a byproduct, yields the list of cluster galaxy members.
Here we use $100$ galaxy clusters with mass $M_{200}\ge 10^{14}h^{-1} M_\odot$ extracted
from a cosmological $N$-body simulation of a $\Lambda$CDM universe to test
the ability of the caustic technique to identify the cluster galaxy members.
We identify the true three-dimensional members as the gravitationally bound galaxies. 
The caustic technique uses the caustic location in the redshift
diagram to separate the cluster members from the interlopers. 
We apply the technique to mock catalogs containing 1000 galaxies 
in the field of view of $12 h^{-1}$~Mpc on a side
at the cluster location. On average, this sample size roughly corresponds to 180 real galaxy members
within $3r_{200}$, similar to recent redshift surveys of cluster regions. 
The caustic technique yields a completeness,
the fraction of identified true members, $f_c=0.95\pm 0.03$ within $3r_{200}$.
The contamination, the fraction of interlopers in the observed catalog of members,  
increases from $f_i=0.020^{+0.046}_{-0.015}$ at $r_{200}$ to 
$f_i=0.08^{+0.11}_{-0.05}$ at $3r_{200}$. 
No other technique for the identification of the
members of a galaxy cluster provides such large completeness and 
small contamination at these large radii.
The caustic technique assumes spherical symmetry and the asphericity of
the cluster is responsible for most of the spread of the completeness and the contamination.
By applying the technique to an approximately spherical system obtained by stacking
the individual clusters, the spreads decrease by at least a factor of two.
We finally estimate the cluster mass within $3r_{200}$ after removing the interlopers:
for individual clusters, the mass estimated with the virial theorem is unbiased and within 30\%
of the actual mass; this spread decreases to less than 10\% for 
the spherically symmetric stacked cluster.

\end{abstract}


\section{Introduction}
Galaxy clusters provide crucial information to 
our understanding of the large-scale cosmic structure and to 
constrain cosmological models. They
populate the high-mass tail of the mass function of virialized galaxy systems;
their abundance and redshift distribution depend on the
average density of the universe and the normalization
of the power spectrum of the initial density perturbations 
\citep[e.g.,][]{voit05, diaferio08, borgani08}.
Clusters are a hostile environment to galaxies
and are thus also a unique tool to investigate
the connection between environment and galaxy properties  
\citep[e.g.,][]{mardom01, mar08, ski09,hue09}.  

Separating the galaxies that do actually belong to the cluster from 
the interlopers -the galaxies that happen to lie in the field of view 
but are not dynamically linked to the cluster- is
crucial to derive accurate estimates of the cluster properties, including
its mass \citep{per90}, or the color and star formation gradients 
of its galaxy population \citep{diaferio01}. 

Interloper rejection techniques are numerous and their sophistication
has progressively increased over the years, thanks to the increased
quality and richness of the observational data: over the last decade, the handful of clusters with tens of 
measured redshifts within $\sim 1-2$\Mpc\ of the cluster center has 
increased by at least a factor of 10 \citep[e.g.,][]{rines03, rines06a, geller11}.

Early observations of galaxy clusters do not usually extend into the
outer regions of the system. Early interloper rejection techniques 
identify galaxy members solely on the basis of their redshift
separation from the cluster center. The gravitational potential
well can however become substantially shallower at increasing radius and the 
combination of velocity and radial distance is now an
essential ingredient for the identification of galaxy members
in samples that extend to the cluster virial radius and beyond.

The caustic technique \citep{diaferio97, diaf99, diaf09, ser10} 
identifies the escape velocity profile
of galaxy clusters from their center to radii as large
as $3r_{200}$, where $r_{200}$ is the radius of the sphere
whose average density is 200 times the critical density 
of the Universe. The technique was thus applied
to estimate the gravitational potential well and the mass profiles
of galaxy clusters to radii that extend to the cluster infall
region (see reviews in \citealt{diaf09} and \citealt{ser10}). 
Where the cluster is in the appropriate redshift range for weak
lensing mass estimation and a comparison is thus possible, caustic and lensing
masses agree within 30\% at the virial radius (Diaferio et al 2005, Geller et al 2013), whereas 
at smaller and larger radii the two mass estimates show a systematic offset of at most 50\% and 20\% respectively \citep{geller2012}.

Because the technique
measures the escape velocity profile, a byproduct of the caustic procedure
is the identification of interlopers.
Compared to other interloper rejection algorithms the caustic technique 
has two major advantages: (1) it does not require the system to be in dynamical equilibrium 
and (2) it does not rely on the derivation of the cluster mass profile to remove interlopers.
These advantages enable the technique to identify interlopers both
in the central and outer regions of clusters, where other 
techniques can not be applied. The caustic technique assumes spherical symmetry, an assumption that is common to most methods.
In addition, when used as a mass estimator method, the caustic technique returns
correct mass estimates if clusters form by hierarchical clustering and thus they 
have the internal kinematical and dynamical properties, including
the shape of the velocity anisotropy profile, that clusters generally have in these models.

The caustic technique as an interloper rejection algorithm, or
some simplified versions of it, was applied to real clusters
to investigate the dependence of galaxy properties on environment 
\citep[e.g.,][]{rin00,Rines04, Rines05, mahajan09, hernandez12, hwang12},  
and to provide robust estimates of the cluster velocity dispersion and mass 
\citep[e.g.,][]{benatov06, lem09, zan11, zhang12}.

Thanks to the approximate self-similarity of self-gravitating systems, 
the technique can also be applied to reject stellar interlopers in galaxies:
\citet{brown09} used the caustic method results to estimate 
the velocity dispersion profile of the stars in the Milky Way halo, and
\citet{ser09} demonstrated that a proper stellar interloper rejection
alleviates the tension between the internal velocity dispersion profiles 
of the Milky Way dwarf satellites and the expectations of Modified Newtonian Dynamics. 
\cite{yegorova11} also probed the dark matter distribution
in the outer regions of disk galaxies by identifying their 
satellites with the caustic technique.

Despite this extensive application, the caustic technique has never been 
exhaustively explored as a method to identify interlopers.
Here, we provide a thorough analysis of its performance and of its random
and systematic errors. In Section \ref{sec:review} we briefly
describe the caustic technique, whereas in Section \ref{sec:cat} we present the mock
cluster catalogs. In Section \ref{sec:idmem} we discuss the technique performance.
We finally investigate the impact of our interloper rejection 
on the cluster mass estimates in Section \ref{sec:massestID}. We compare the performance of
our method with other rejection techniques in Section \ref{sec:disID}. Conclusions are presented in Section \ref{sec:conID}.

\section{The Caustic Technique}
\label{sec:review}

In hierarchical clustering, clusters of galaxies form by the
aggregation of smaller systems. The accretion is not purely radial \citep[e.g.,][]{whi10}, 
because galaxies within the
falling clumps have velocities with a substantial non-radial component.
Therefore, the galaxy velocities are set by the 
local gravitational potential more than by the radial infall expected 
in the spherical collapse model \citep{diaferio97}.

When observed in the  redshift diagram -the
plane of the line-of-sight velocity $v$ of the galaxies in the cluster rest frame 
versus their projected distance $r$ from the cluster center-
the cluster members populate a region with a trumpet shape approximately
symmetric along the $r$ axis \citep{kaiser87,regos89,vanhaarlem93}.
The caustics define the boundaries of this region whose amplitude 
${\cal A}(r)$ decreases with increasing $r$. 
\cite{diaferio97} demonstrate 
that ${\cal A}(r)$ is a combination of the profile of the escape velocity from
the cluster and the profile of the velocity anisotropy parameter $\beta(r)=1-(\langle
v^2_\theta\rangle + \langle v^2_\phi\rangle)/2\langle v^2_r\rangle $, where
$v_\theta$, $v_\phi$, and $v_r$ are the longitudinal, azimuthal and
radial components of the velocity ${\mathbf v}$ of a galaxy, respectively, and the
brackets indicate an average over the velocities of the galaxies in the volume
$\mbox{d}^3{\bf r}$ centered on position ${\bf r}$. 

In a spherically symmetric system, the average square of the velocity
of the system members at radius $r$ is $\langle
v^2\rangle=\langle v^2_{\rm los} \rangle g(\beta)$ where
$\langle v^2_{\rm los} \rangle$ is the component of the line-of-sight  
velocity and 
\begin{equation}
g(\beta) = {3-2\beta(r)\over 1-\beta(r)}\; . 
\end{equation}

For the escape velocity at radius $r$, we have $\langle
v_{\rm esc}^2(r)\rangle=-2\phi(r)$, where $\phi(r)$ is the gravitational potential. 
If the amplitude ${\cal A}(r)$ measures the average
component along the line of sight of the escape velocity at radius
$r$, namely ${\cal A}^2(r)=\langle v^2_{\rm esc, los}\rangle$, we obtain the relation 
\begin{equation}
-2\phi(r)={\cal A}^2(r)g(\beta) \equiv \phi_\beta(r) g(\beta) \; .
\label{eq:rig-pot}
\end{equation}

This equation shows the dynamical information contained in the observable caustic amplitude 
${\cal A}(r)$. Being a combination of the gravitational 
potential profile and the function $g(\beta)$, ${\cal A}(r)$ can provide the 
estimate of both the escape velocity profile from the cluster and the mass profile of the cluster \citep{diaferio97, diaf99}. We emphasize that the entire argument outlined above
holds regardless of the stability of the system. 

To measure ${\cal A}(r)$ we need to locate the 
caustics in the redshift diagram. 
The technique consists of three major steps: (1) the construction
of a binary tree based on the projected galaxy pairwise energy; (2) the
determination of a threshold to cut
the binary tree; (3) the identification of the cluster center to 
obtain the redshift diagram and determine the galaxy number density
on this diagram.

\begin{figure*}
\includegraphics[angle=0,scale=.85,bb=70 90 680
550]{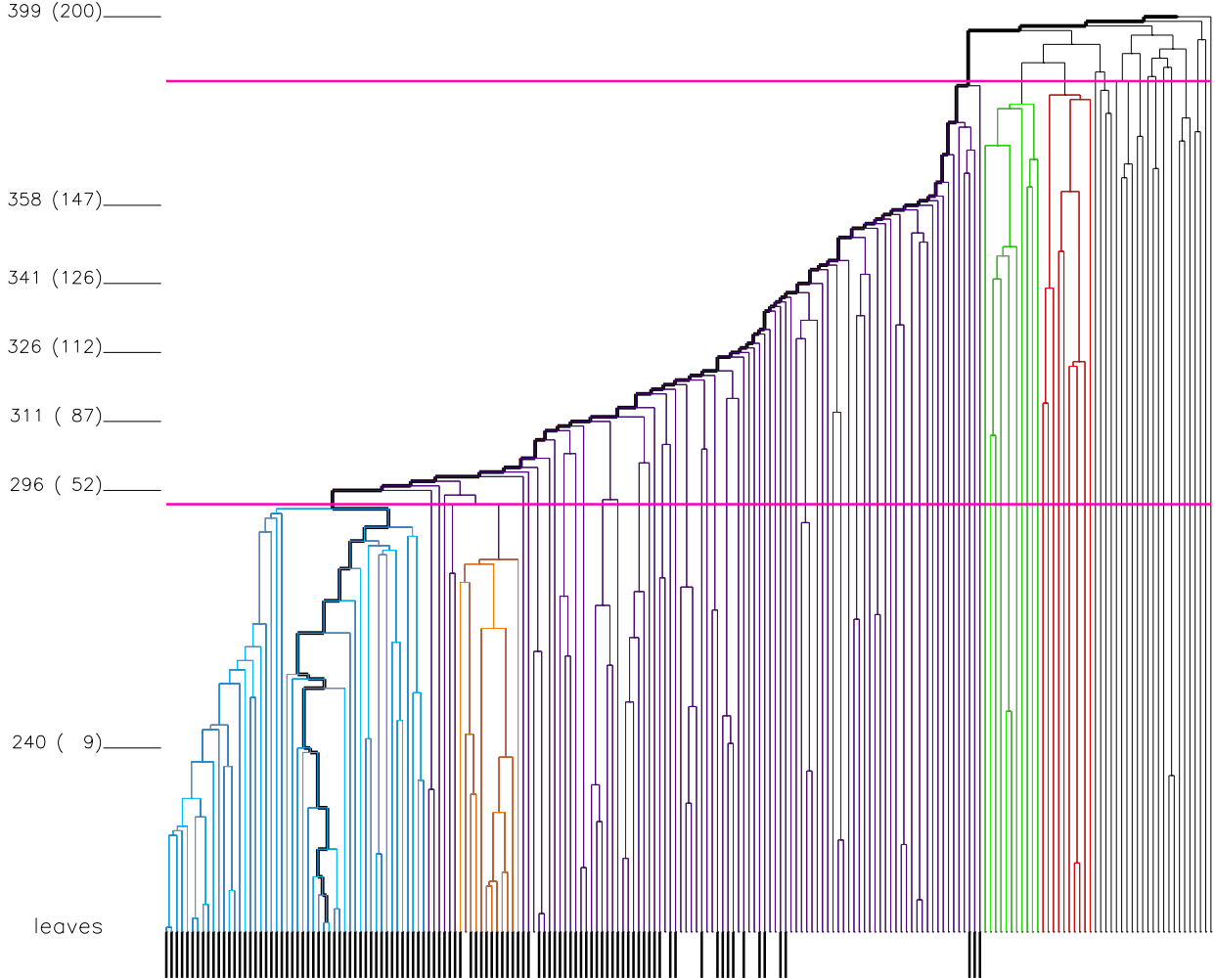}
\caption{Dendrogram representation of the binary tree of a random subsample of 200
particles in the field of view of a simulated cluster. 
The particles are the leaves of the tree at the bottom of the plot. The
particles within $3r_{200}$ in real space are highlighted in black. The thick path
highlights the main branch of the tree. The horizontal lines show
the levels at the two nodes $x_1$ (upper line) and $x_2$ (lower line)
that limit the $\sigma$ plateau shown in Figure \ref{fig:plateau}. The upper
node $x_1$ is the threshold where the tree is cut and the main group is the structure
hanging from this node. 
Only as a guide, some nodes are labeled on the left-hand side, with
their number of associated particles, the descendants, in brackets.} 
\label{fig:dendrogram}
\end{figure*}

\begin{figure}
\includegraphics[angle=0,scale=.53,bb= 100 14 566
385]{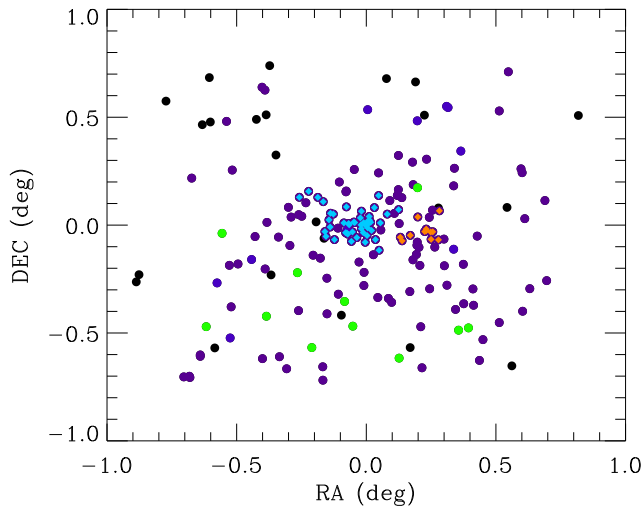}
\caption{Celestial coordinates of the subsample of 200 particles in the field of 
view of the simulated cluster whose binary tree is shown in 
Figure \ref{fig:dendrogram}. The color code is the same as in Figure
\ref{fig:dendrogram}.}
\label{fig:sky}
\end{figure}

\begin{figure}
\includegraphics[angle=0,scale=.50,bb=60 14 566 390]
{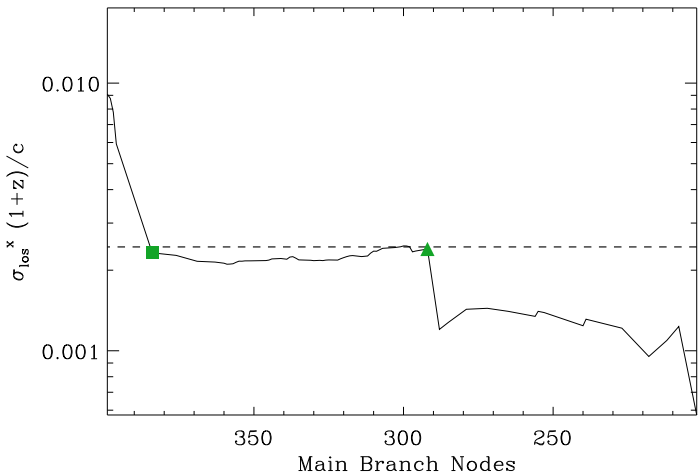}
\caption{Velocity dispersion of the leaves of each node
along the main branch of the binary tree shown in Figure \ref{fig:dendrogram}. 
The square and the triangle show the nodes $x_1$ and $x_2$ respectively. 
The curve between $x_1$ and $x_2$ is
the $\sigma$ plateau. The dashed line shows the line-of-sight velocity 
dispersion of the particles within the sphere of radius $3r_{200}$.}
\label{fig:plateau}
\end{figure}

At the first step, all the galaxies are arranged in a
binary tree according to their pairwise binding energy
\begin{equation}
E_{ij}=-G{m_i m_j\over R_p}+{1\over 2}{m_i m_j\over m_i+m_j}\Pi^2 \;,
\label{eq:pairwise-energy}
\end{equation}
where $R_p$ is the pair projected separation, $\Pi$ is the line-of-sight velocity difference and $m_i=m_j=10^{12}h^{-1}$ M$_\odot$ are the two galaxy masses assumed to be constant.

The binary tree is built as follows: (i) initially each galaxy is a group
$g_\alpha$; (ii) the binding energy
$E_{\alpha\beta}={\rm min}\{E_{ij}\}$, where $E_{ij}$ 
is the binding energy between the galaxy $i\in g_\alpha$ and the
galaxy $j\in g_\beta$, is associated with each
group pair $g_\alpha, g_\beta$;
(iii) the two groups with the smallest binding energy $E_{\alpha\beta}$ are
replaced with
a single group $g_\gamma$ and the total number of groups is decreased by one;
(iv) the
procedure is repeated from step (ii) until only one group is left. Figure
\ref{fig:dendrogram} shows the binary tree of a random sample of 200  
particles extracted from a simulated
halo selected from the $N$-body simulation described in the 
next section, whereas Figure \ref{fig:sky} shows the celestial coordinates of the same
particles with the same color code as in Figure \ref{fig:dendrogram}.

The second step of the caustic technique procedure
is the threshold choice. The tree 
arranges the galaxies in potentially distinct groups; however, 
to get effectively distinct groups and to specifically define the set of
candidate members, we need to cut the tree at some level. This level sets the node from
which the candidate members hang. All these candidate members
do not necessarily coincide with the optimal members that are 
determined by the caustic location. Below we will
extensively illustrate the reason for this
distinction between candidate and optimal members. 

In order to choose the threshold to cut the
binary tree,  we identify the main
branch as the branch that emerges from the 
root and contains the nodes from which, at each level, the largest number of
galaxies (or leaves) hangs. The leaves
hanging from each node $x$ of the main branch provide a velocity dispersion
$\sigma_{\rm{los}}^x$.
When walking along the main branch from the root to the leaves, $\sigma_{\rm{los}}^x$  
rapidly decreases due to the progressive loss of galaxies that are most likely not
associated with the cluster (Figure \ref{fig:plateau}); then $\sigma_{\rm los}^x$ reaches a ``$\sigma$
plateau'' at some node $x_1$. Most of the galaxies hanging from this node are members;
in fact, the system is nearly isothermal and the removal of the less bound
galaxies does not affect the value of $\sigma_{\rm{los}}^x$. At some point of
the walk along the main branch, the loss of the most bound galaxies, whose binding energy is
very small, causes $\sigma_{\rm{los}}^x$ to drop again. This second rapid drop 
identifies the nodes $x_2$ which sets the limit of the $\sigma$ plateau. The first node $x_1$ 
closest to the root is the appropriate level for the identification of the system and
we define the galaxies hanging from it the candidate members of the cluster. They 
determine the center of the system, its radius, and its line-of-sight velocity dispersion.
These quantities are used to build the redshift diagram.

The  third step of the procedure is the location of the caustics in the redshift diagram.
The caustics are the curves satisfying the equation $f_q(r,v)=\kappa$. Here
$f_q(r,v)$ is the galaxy number density in the redshift diagram, namely
the plane $(r,v)$, and $\kappa$ is the root of the equation
\begin{center}
\begin{equation}
\langle v_{\rm esc}^2\rangle_{\kappa,R}=4\langle v^2\rangle\; .
\label{eq:skappa}
\end{equation}
\end{center}
The function $\langle v_{\rm esc}^2\rangle_{\kappa,R}=\int_0^R{\cal
A}_\kappa^2(r)\varphi(r)dr/
\int_0^R\varphi(r)dr$ is the mean caustic amplitude within $R$,
$\varphi(r)=\int f_q(r,v) dv$, $\langle v^2\rangle^{1/2}$ is
the velocity dispersion of the candidate members, 
$R$ is their mean projected separation from the center, and
$q$ is a smoothing parameter (see \citealt{diaferio97, ser10} for details). 

\begin{figure}
\includegraphics[angle=0,scale=.53,bb= 100 14 566
395]{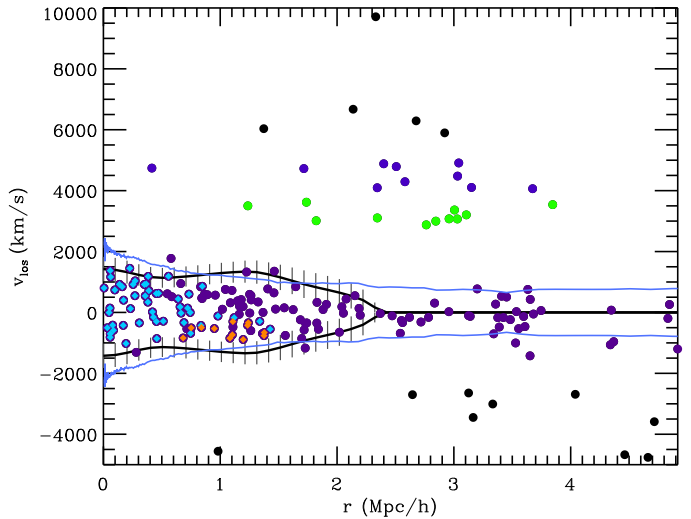}
\caption{Redshift diagram of the subsample of 200 particles in the field of view of the
simulated cluster whose binary tree is shown in Figure \ref{fig:dendrogram}. 
The black lines with error bars 
are the caustics located by the caustic technique. The cyan lines are the
real caustics determined by the profiles of the escape velocity 
and the velocity anisotropy parameter derived by the three-dimensional information. 
The dots show
the particles in the catalog and the color code is as in Figure \ref{fig:dendrogram}.}
\label{fig:rediag}
\end{figure}

Figure \ref{fig:rediag} shows the result of this procedure on
the redshift diagram. We stress that the procedure to locate the caustics is independent
of any assumption on the dynamical equilibrium of the system, 
of the shape of $g(\beta)$ and of the gravitational potential profile $\phi(r)$;
this procedure actually measures the combination of $g(\beta)$ and $\phi(r)$ 
expressed by the caustic amplitude ${\cal A}(r)$ (Equation \ref{eq:rig-pot}).

\section{Simulated clusters and mock catalogs}

\label{sec:cat}

We use the synthetic galaxy clusters described in \citet{ser10}
selected from the $N$-body simulation of \citet{borgani04}. 
The simulation models a cubic volume of 192~$h^{-1}$~Mpc 
on a side of a flat $\Lambda$CDM model, with matter density
$\Omega_0=0.3$, Hubble parameter $h=0.7$, 
normalization of the power spectrum $\sigma_8=0.8$ and
baryon density $\Omega_{\rm b}=0.02h^{-2}$. The simulation contains
$480^3$ dark matter particles with mass $m_{\rm DM}=4.6\times 10^9h^{-1}$~M$_\odot$
and, initially, $480^3$ gas particles with mass $m_{\rm gas}=6.9\times 10^8h^{-1}$~M$_\odot$. 
The simulation was run with GADGET-2 \citep{springel2005}. Further details
of the simulations and the dark matter halo identification 
are given in \citet{borgani04}. In the following, we limit our analysis to the gravitational
dynamics of the dark matter distribution. In fact, both $N$-body simulations
\citep[e.g.,][]{diaf01, gill04, diemand04,
gill05} and observations \citep[e.g.,][]{rines08}
indicate that any velocity bias between galaxies and dark matter is negligible. 

We consider the 100 dark matter halos with mass
$M(<r_{200})\equiv M_{200} \ge 10^{14} h^{-1} M_\odot$ at redshift $z=0$.
We locate each halo at $(\alpha,\delta)=(6^h,0^\circ)$ and redshift $cz=32000$ km~s$^{-1}$.
We simulate the compilation of the redshift catalog of a galaxy cluster 
by projecting each halo along 10 random
lines of sight. For each of these lines of sight, we choose two additional
directions orthogonal to the first one and to each other. We end up with $3000$ mock redshift catalogs. 
Each catalog contains a random sample of $1000$ particles distributed
within a rectangular parallelepiped centered on the cluster with a squared 
field of view of $12 h^{-1}$~Mpc on a side and 192~$h^{-1}$~Mpc deep.
With this number of particles in the field of view, we obtain
a distribution of the number of particles within the sphere of radius
$r_{200}$ that has median $101$ and percentile range [10\%, 90\%] equal to $[51,226]$;
the median number of particles within $3r_{200}$ is 185 and the  percentile range [10\%, 90\%] 
is $[96, 408]$. These numbers are comparable to the sample sizes
of recent large galaxy redshift surveys of
clusters and their surroundings, such as CIRS \citep{rines06a} and HeCS \citep{rines12}.

The binary tree algorithm applied to the individual mock catalogs gives a center 
of the cluster and a velocity dispersion of the candidate members. 
The center and velocity dispersion determined with the binary tree are close to 
the correct quantities in most cases \citep{ser10}. Specifically, in 2678 mock catalogs
(89\% of the cases) the algorithm locates the center on the
expected cluster; in the remaining 11\% of the cases, the 
field of view is particularly crowded with numerous groups and clusters,
and the cluster of interest might not be the most massive cluster in the field. In these cases, 
the algorithm identifies the center of a different cluster. In a similar
situation happening with catalogs of real clusters we will relocate the center on the 
cluster of interest. Here, we simply remove
these problematic catalogs. Among the 2678 correctly identified clusters, the estimated velocity
dispersion within $3r_{200}$ is within 5 (30)\% of the real one in 50 (95)\%
of the systems; the
center deviations are smaller than $0^\circ.07$ on the sky and 250~km~s$^{-1}$ along
the line of sight in 90\% of the clusters. The largest discrepancies
between the correct center and the center found by the algorithm occur in systems
with evident substructures that produce multiple peaks of the particle number density distribution. 
When happening with catalogs of real clusters, these cases can 
yield off-centered redshift diagrams. This problem can be removed by 
relocating the center on the most luminous galaxy of the cluster or on the peak
of the X-ray emission. In our mock catalogs, we do not keep these systems, but further remove 
those catalogs where the center found by the algorithm has an offset greater 
than $0.5$\Mpc\ on the sky or greater than $400$~km~s$^{-1}$ along the
line of sight. Thus, the final number of mock catalogs reduces from 2678 to 2420. 

\section{Identification of cluster members} \label{sec:idmem}

\subsection{Definition of Members} \label{sec:def}

\begin{figure}
\includegraphics[angle=0,scale=.63, bb=65 0 584 340]{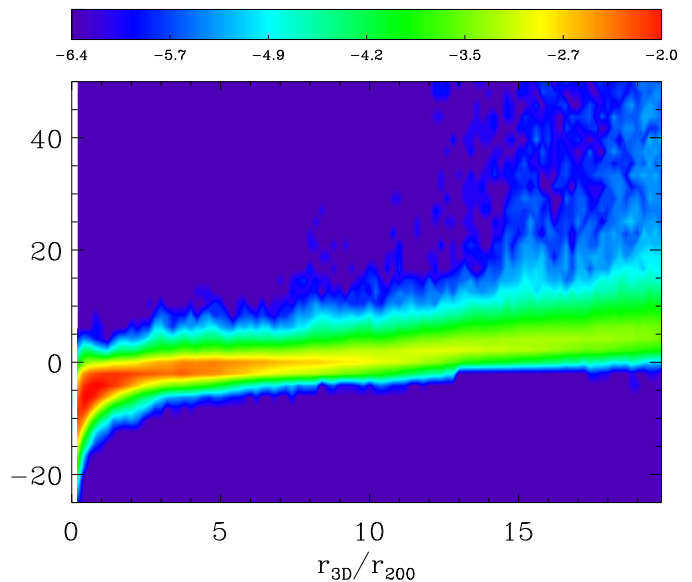}
\caption{Logarithm of the number density distribution of the 
particles in the plane of binding energy versus clustrocentric
distance. The top bar shows the color scale, with the number density distribution  
normalized to the total number of particles in the diagram. Binding energies 
and clustrocentric distances are normalized to the three-dimensional  
velocity dispersion $\sigma$ and $r_{200}$ of each individual cluster, respectively. 
The number density distribution includes all the particles from the entire
sample of 2420 mock catalogs.}
\label{fig:enerdist}
\end{figure}

A galaxy is a cluster member if its
binding energy is negative, namely if its velocity $v$ is lower than the velocity $v_{\rm esc}(r)$
required to escape the cluster when the galaxy is at distance $r$ from the cluster center: 
$v^2<v_{\rm esc}^2(r)$. Figure \ref{fig:enerdist}
shows the number density distribution of the dark matter particles
in our mock cluster catalogs in the plane of binding energy versus the three-dimensional (3D) clustrocentric 
distance. The plot includes the entire sample of 2420 cluster catalogs. 

Figure \ref{fig:enerdist} shows that a substantial fraction of bound particles have
clustrocentric distance much larger than $r_{200}$ \citep{woj08}. Therefore, in principle,
we might use a different and simpler criterion to define a cluster member: a galaxy
whose clustrocentric distance is smaller than, for example, $3r_{200}$. This criterion is
actually more restrictive than the criterion based on the binding energy, as
Figure \ref{fig:enerdist} suggests. Nevertheless, we include this criterion
in the following analysis, for the sake of comparison. 
Hereafter, we call 3D members these two sets of members defined on the basis of 
the 3D information.

We expect that the caustic technique identifies the 3D members 
as the galaxies within the caustics in 
the redshift diagram. In the following we will also consider a criterion 
based on the binary tree. As described in Section \ref{sec:review}, 
the caustic technique arranges the galaxies in a binary tree according to
their pairwise projected binding energy; by cutting the 
tree at the $\sigma$ plateau we define a set of candidate members.
In the following analysis, we show that
the choice of this name is appropriate, because the interloper 
contamination of this set of candidate members is larger than
the contamination of the set of members determined by the caustic location. 

In conclusion, we consider two definitions of 3D members: (a) galaxies
with negative binding energy; (b) galaxies within $3r_{200}$; and 
two possible criteria for their identification: 
(1) galaxies within the caustics in the redshift diagram; (2) 
galaxies on the main branch of the binary tree cut at the $\sigma$ plateau.
Hereafter, we refer to the members identified with methods (1) or (2) as 2D members.

In Sections \ref{sec:causID} and \ref{sec:treeID} we focus on the 
performance of the first and second criteria, respectively.
We compute two relevant quantities: the completeness
$f_c$, which is the fraction of 3D members that are also identified as 2D members, and the
contamination $f_i$, which is the ratio between the number of particles 
taken as 2D members that are actually interlopers and the total number of
2D members.

\subsection{2D Members: Caustic Location} \label{sec:causID}

\begin{figure}
\includegraphics[angle=0,scale=.47,bb= 40 14
566 390]{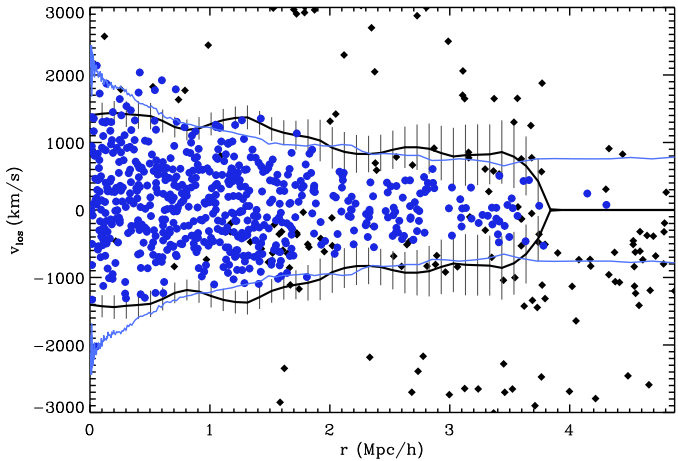}
\caption{Redshift diagram of $1000$ particles of a synthetic cluster. 
The black lines with 1-$\sigma$ error bars and the cyan  
lines are the estimated and true caustics respectively.
The symbols are the particles in the catalog; the blue dots are
the bound particles. There are 636 particles between the upper and lower
caustics. } 

\label{fig:rediagbound}
\end{figure}

\begin{figure*}
\includegraphics[angle=0,scale=.55,trim=0 0 0 0]{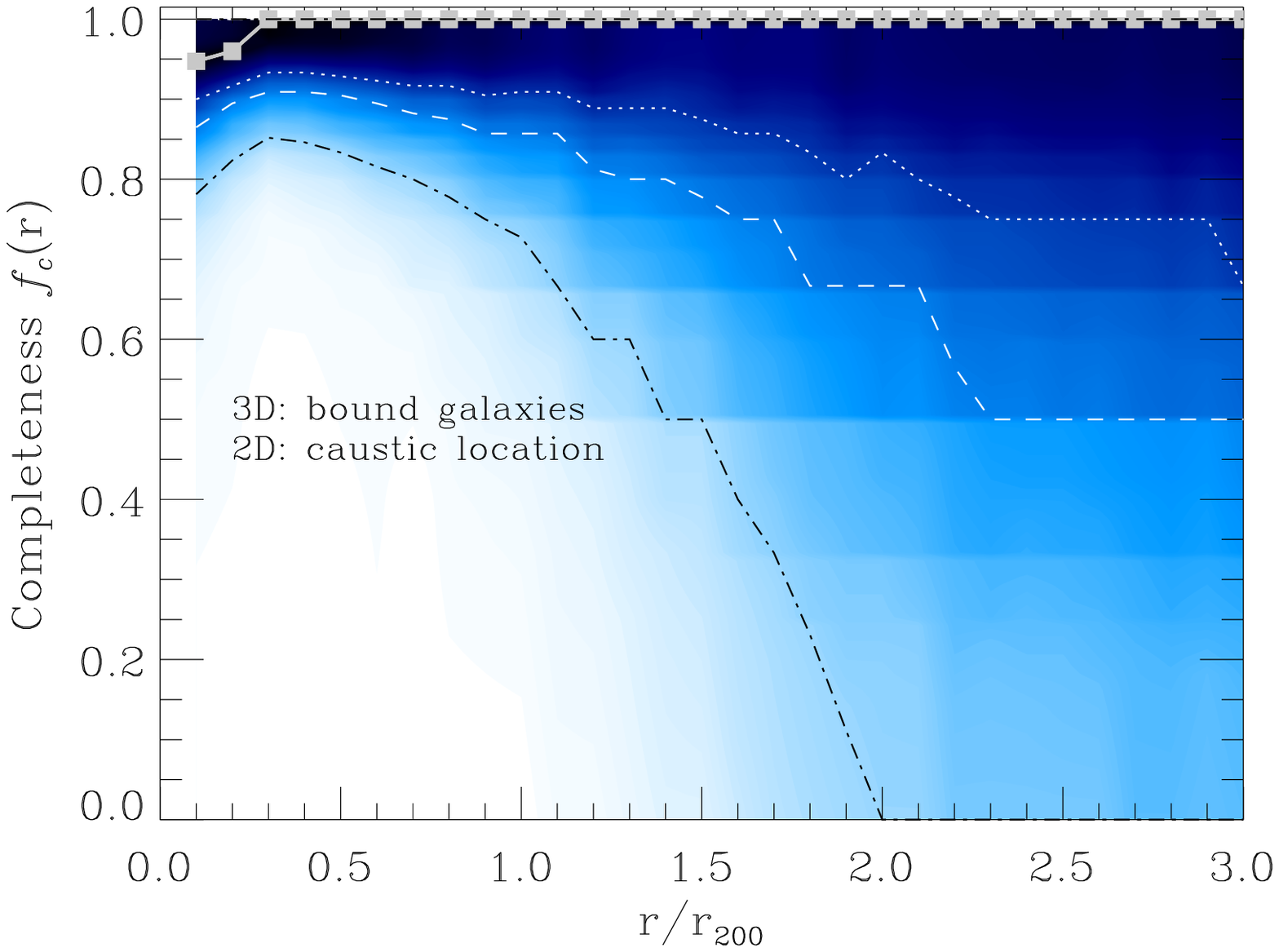}
\includegraphics[angle=0,scale=.55,trim=0 0 0 0]{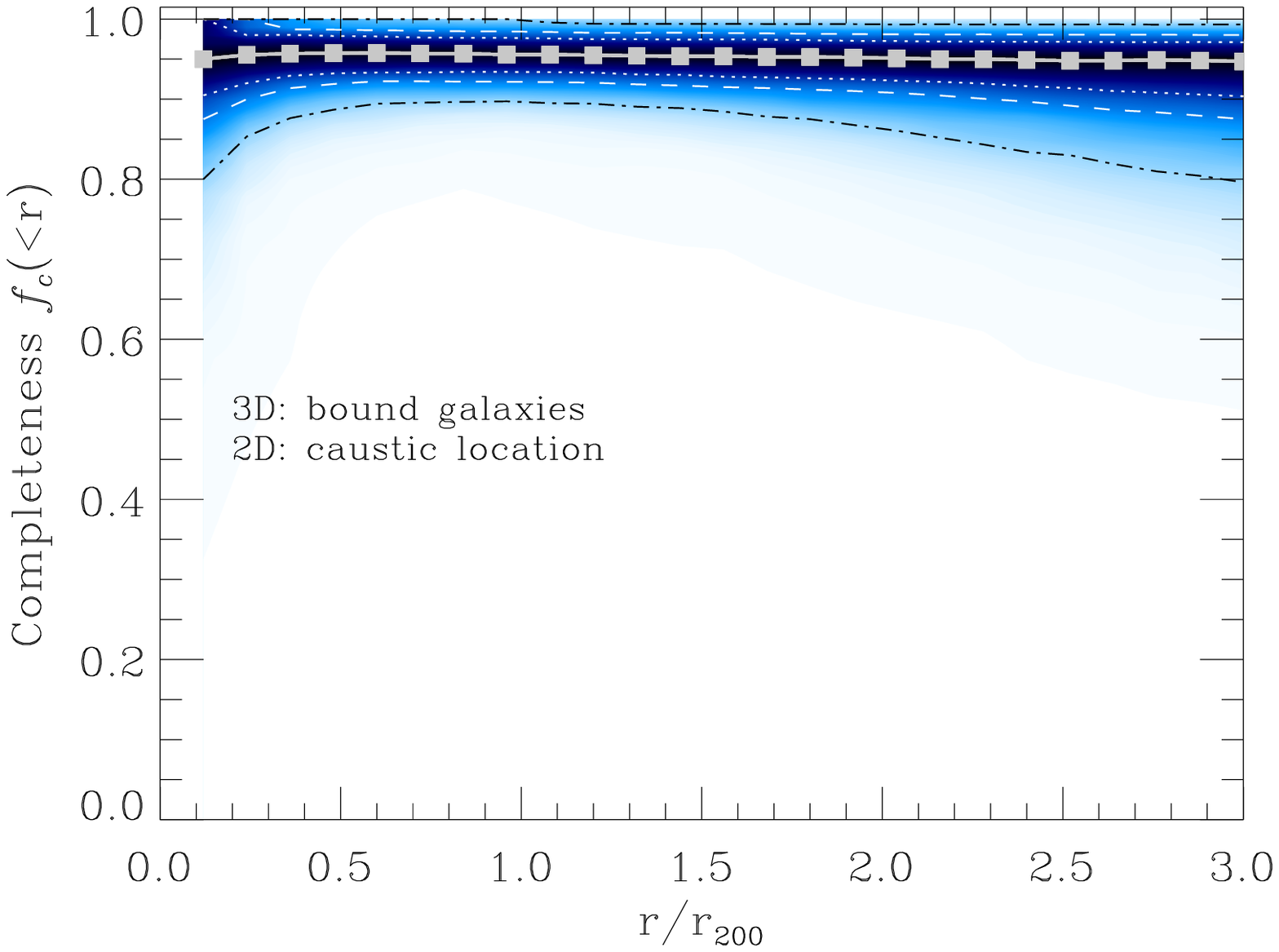}
\includegraphics[angle=0,scale=.55,trim=0 0 0 0]{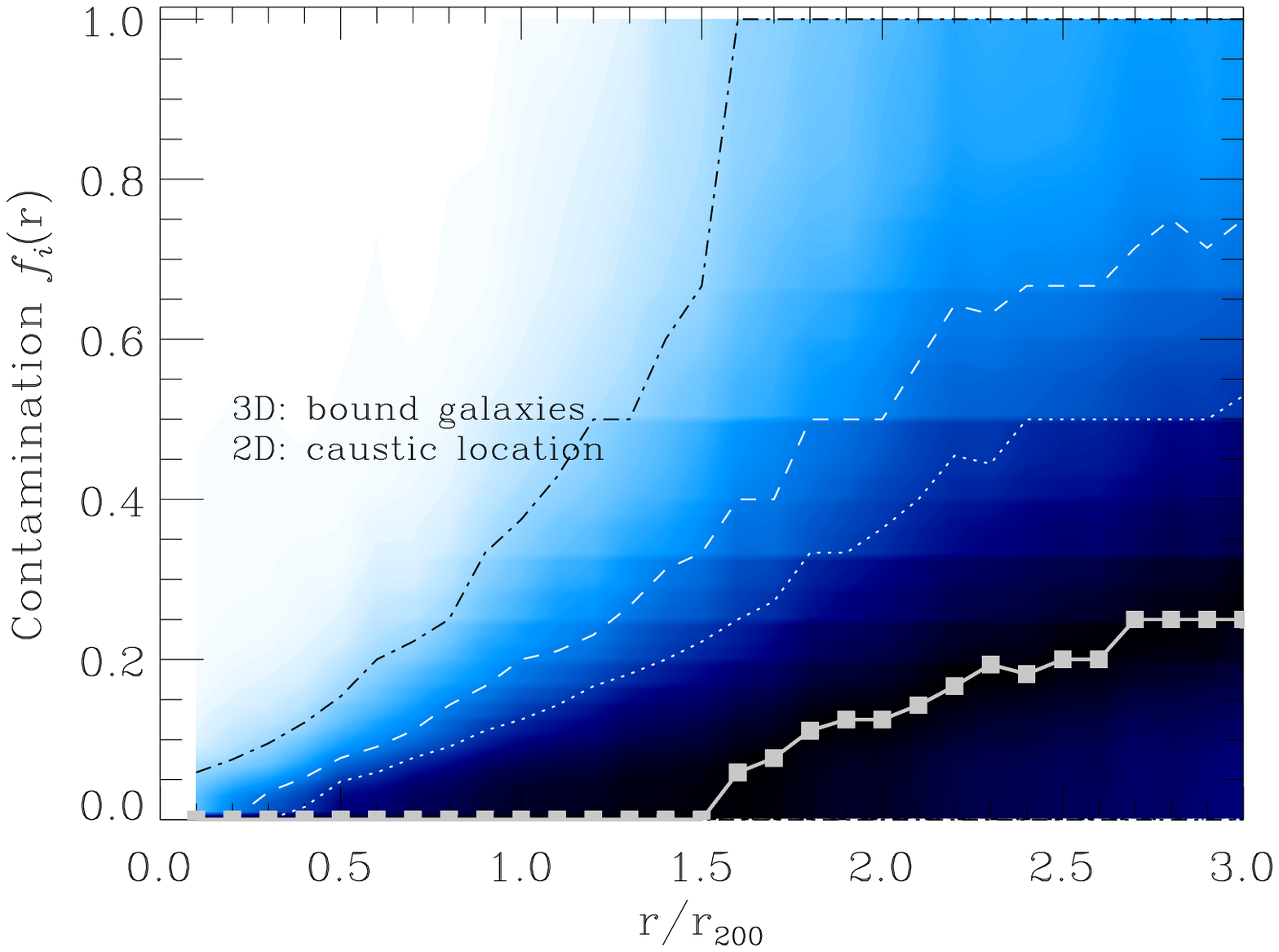}
\includegraphics[angle=0,scale=.55,trim=0 0 0 0]{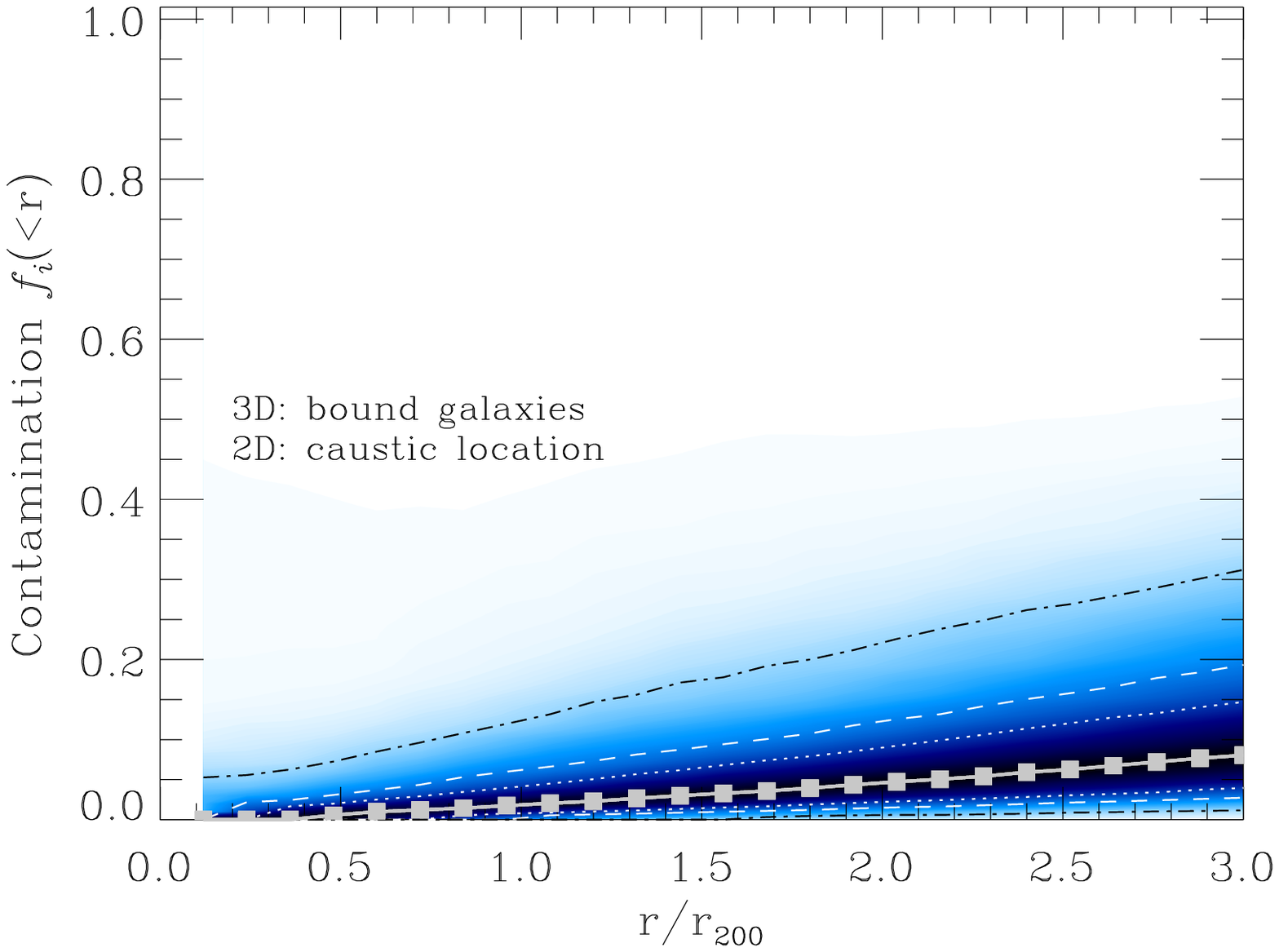}
\caption{Upper panels: differential (left panel) and cumulative (right panel) median profiles (solid squares) of the completeness $f_c$, where the 3D members are the bound galaxies and the 2D members are the galaxies within the caustics. 
Lower panels: differential (left panel) and cumulative (right panel) profiles of the contamination $f_i$. The darkness of the shaded areas is proportional
to the profile number density on the vertical axis. The dotted, dashed, and dot-dashed lines limit
the areas including the 50\%, 68\%, and 90\% of the profiles, respectively.}
\label{fig:evel-fg}
\end{figure*}

Figure \ref{fig:rediagbound} shows the redshift diagram of a cluster
from our sample, with the bound galaxies defined as the 3D members, 
shown as blue dots. As expected, most of the 3D members are within the caustics. 

To illustrate how the method performs on average in this case, we compute the completeness
and contamination profiles of each cluster. At each radius, we consider the median
of the set of profiles and their dispersion. 
The upper left panel of Figure \ref{fig:evel-fg} shows the median differential profile of the
completeness $f_c$ and the regions containing 50\%, 68\% and 90\% of
the profiles.

Only at small radii the caustic algorithm 
removes a few per cent of the 3D members, because the caustic amplitude
is slightly underestimated, as can be seen in the example of Figure \ref{fig:rediagbound}.
The caustic criterion thus provides a completeness close to 0.95 at radii
smaller than $0.2r_{200}$, and increases to 1.0 at larger radii.

The median differential contamination (Figure \ref{fig:evel-fg}, bottom left panel) is larger than 0.1 at radii larger than $2r_{200}$, but the cumulative contamination (Figure \ref{fig:evel-fg}, bottom right panel) remains below 0.09 at $3r_{200}$. 
Table \ref{table:comp} gives the corresponding 68\% levels 
for the cumulative profiles of $f_c$ and $f_i$.

\begin{figure}
\includegraphics[angle=0,scale=.47,bb= 40 14 566 390]{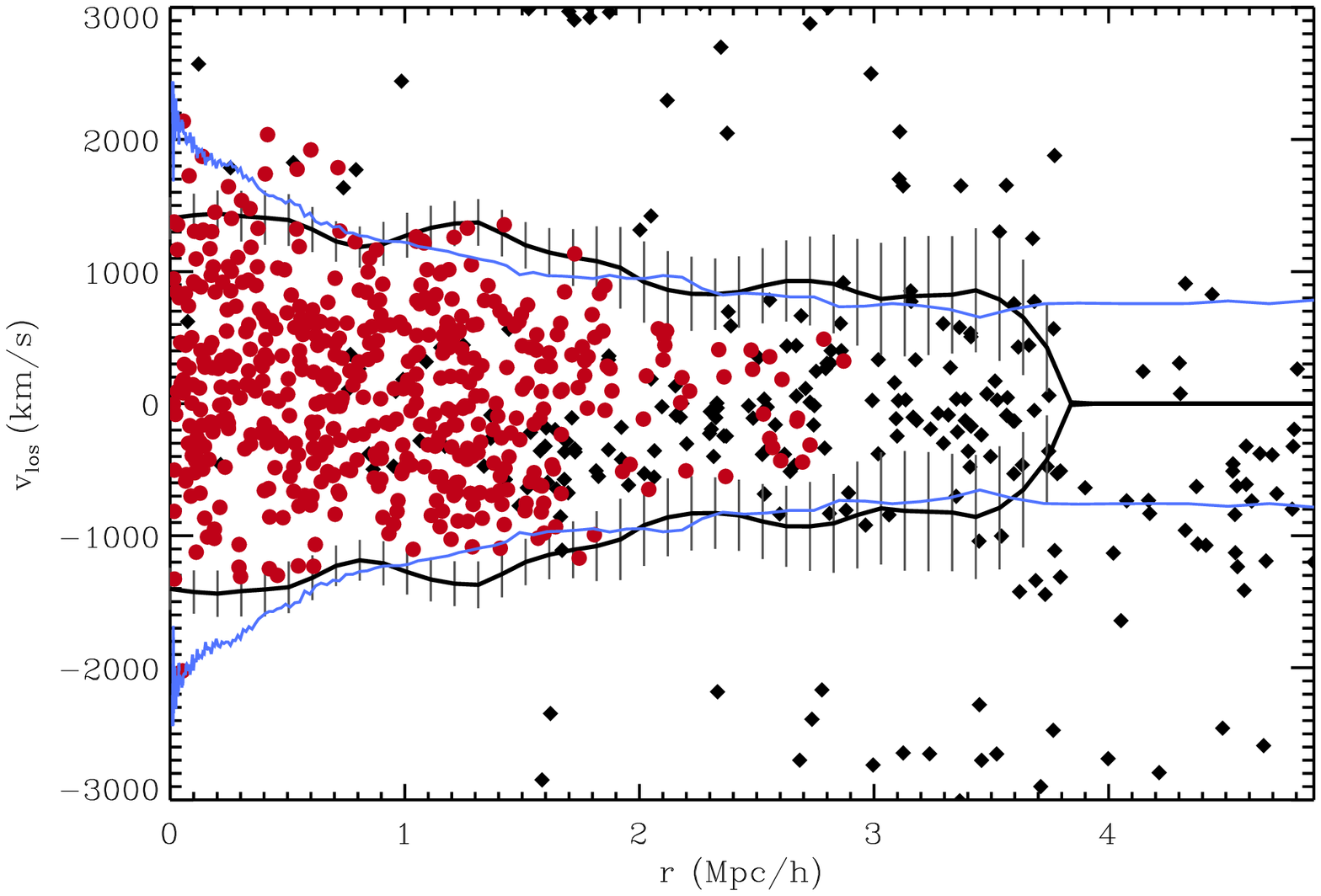}
\caption{Same as Figure \ref{fig:rediagbound}, with red dots
the 3D members defined as the particles within the sphere of radius $3r_{200}$.}
\label{fig:rediag3r200}
\end{figure}

Figure \ref{fig:rediag3r200} reproduces the same redshift diagram 
of Figure \ref{fig:rediagbound} with the red dots the particles within
$3r_{200}$ from the cluster center defined as the 3D members. In this case, 
most 3D members are within the caustics, but, at large radii, many particles within 
the caustics are not 3D members. In fact, in a sample of particles extracted 
from a spherical halo whose number
density profile decreases with radius, the number of particles
within a given 3D radius $r$ observed in projection can fall to zero
beyond a projected radius $r_e\le r$; obviously, $r_e$ decreases with the
size of the particle sample. For a \citet*{NFW} number density profile
and for a sample of $185$ particles within $3r_{200}$ (the average number of particles per cluster sample),
we find that the number of particles with a 3D distance smaller than
$3r_{200}$ is zero at projected distances larger than $r_e \sim 2.6 r_{200}$. The differential contamination $f_i$ for the entire sample
remains smaller than 0.1 within $r_{200}$, on average, but increases dramatically
at larger radii and reaches $f_i=1$ at $r\sim 2.6 r_{200}$. 
It follows that the corresponding cumulative profile 
of the fraction of galaxies identified as members when they are actually interlopers 
reaches 0.27 at $3r_{200}$ (Table \ref{table:comp}). 

On the other hand, the profiles of the completeness we obtain with this definition of
3D members are very similar to Figure \ref{fig:evel-fg}:
the method yields a large ($\sim 0.95$) and stable completeness to radii as large
as $3r_{200}$ (Table \ref{table:comp}).

\subsection{2D Members: Main Group of the Binary Tree}\label{sec:treeID}

We now evaluate the completeness $f_c$ and the contamination $f_i$ of the set of members identified with the binary tree.

With the bound galaxies as the 3D members, the completeness $f_c=1.00$ is  
constant over the entire range of $r$ (Table \ref{table:comp}). 
The median differential contamination $f_i$
reaches a maximum value of $\sim0.4$ at $r>2.5r_{200}$. The
cumulative profile thus reaches the value 0.13 (Table \ref{table:comp}) at $r=3r_{200}$. 
The differential profile of $f_i$ shows that, if the bound galaxies are 
taken as 3D members, using the binary tree
procedure introduces interlopers at $r>0.8r_{200}$; this result
is somewhat worse than the caustic location performance shown in the previous section, because 
the caustic location, on average, provides
samples without contamination up to $r=1.5r_{200}$ (Figure \ref{fig:evel-fg}). Overall, 
however, when we adopt the bound particles as 3D member, both 
the caustic location and the binary tree give high levels of completeness ($f_c\sim 0.95-1.0$) and low levels of contamination ($f_i\sim 0.08-0.13$) within $3r_{200}$.

In the case of the galaxies within
$3r_{200}$ as 3D members (definition (b) in Section \ref{sec:def}), the completeness $f_c$ has a constant median
value $f_c=1.0$ (Table
\ref{table:comp}). Clearly, applying the binary tree procedure to determine the members of a cluster
guarantees an extremely high completeness of the sample. On the other hand, the median differential 
contamination $f_i$ 
is smaller than 0.1 at $r<r_{200}$, and increases at larger radii up to $1$. This high
contamination at large radii translates into a cumulative $f_i$ of 0.31 at $3r_{200}$ (Table \ref{table:comp}). The reason for this large contamination at large radii derives from 
the decreasing of the number density profile, as discussed
in Section \ref{sec:causID}.

\begin{table*}
\begin{center}

\begin{tabular}{ccccc|ccc||ccc|ccc}
&     &\multicolumn{6}{c}{\textbf{bound galaxies}} &       \multicolumn{6}{c}{\textbf{galaxies within $3r_{200}$}}\\
\textbf{Caustic}&     &      &      &    &\\
\textbf{location} &     &   $f_c-\sigma$ & $f_c$  & $f_c+\sigma$     & $f_i-\sigma$ & $f_i$ & $f_i+\sigma$
      &  $f_c-\sigma$ & $f_c$ & $f_c+\sigma$  &   $f_i-\sigma$ & $f_i$ & $f_i+\sigma$\\
\cmidrule[0.05em](l ){3 -14}

&$r_{200}$    &  0.921  &  0.956  &  0.984  &  0.005  &  0.020  &  0.066
              &  0.917  &  0.953  &  0.983  &  0.014  &  0.042  &  0.118 \\
&$2r_{200}$   &  0.908  &  0.951  &  0.981  &  0.015  &  0.047  &  0.126
              &  0.903  &  0.947  &  0.980  &  0.053  &  0.125  &  0.256 \\
&$3r_{200}$   &  0.875  &  0.947  &  0.980  &  0.027  &  0.080  &  0.193
              &  0.898  &  0.946  &  0.980  &  0.143  &  0.273  &  0.418 \\

\\
\end{tabular}

\begin{tabular}{ccccc|ccc||ccc|ccc}

\textbf{Binary}&     &       &      & &\\
\textbf{tree} &     &   $f_c-\sigma$ & $f_c$  & $f_c+\sigma$     & $f_i-\sigma$ & $f_i$ & $f_i+\sigma$
      &  $f_c-\sigma$ & $f_c$ & $f_c+\sigma$  &   $f_i-\sigma$ & $f_i$ & $f_i+\sigma$\\
\cmidrule[0.05em](l ){3 -14}
&$r_{200}$    &  0.990  &  1.000   &  1.000  & 0.010  &  0.031  &  0.083
              &  0.990  &  1.000   &  1.000  & 0.020  &  0.050  &  0.129 \\
&$2r_{200}$   &  0.981  &  1.000   &  1.000  & 0.034  &  0.078  &  0.156
              &  0.984  &  1.000   &  1.000  & 0.073  &  0.149  &  0.274 \\
&$3r_{200}$   &  0.953  &  0.997   &  1.000  & 0.067  &  0.133  &  0.233
              &  0.980  &  1.000   &  1.000  & 0.190  &  0.306  &  0.443 \\

\\
\end{tabular}
\end{center}
\caption{Cumulative completeness $f_c$ and contamination $f_i$ at $r_{200}$, $2r_{200}$, and $3r_{200}$, and their 1-$\sigma$ dispersion,
where the 3D members are the bound particles (left) or the particles within $3r_{200}$ (right) and the 2D members are the particles within the caustics (top) or in the main group of the binary tree
(bottom).}
\label{table:comp}
\end{table*}

\subsection{Identification of Members in Stacked Clusters} \label{sec:stacked}

As expected, the results listed in Table \ref{table:comp} indicate 
that the caustic location is more effective than the binary tree
algorithm in identifying 3D members and that the binding energy criterion is more appropriate 
than the geometrical criterion to
define members based on 3D data. Table \ref{table:comp} also 
shows the spreads of the completeness and contamination. These spreads  
originate from the random and systematic errors
of the caustic technique, which are mostly due to the assumption of 
spherical symmetry \citep{ser10}. 
In addition, we expect that the 
performance of the technique depends on the number of galaxies in the 
catalog. 

\begin{figure*}
\includegraphics[angle=0,scale=.55,trim=0 0 0 0]{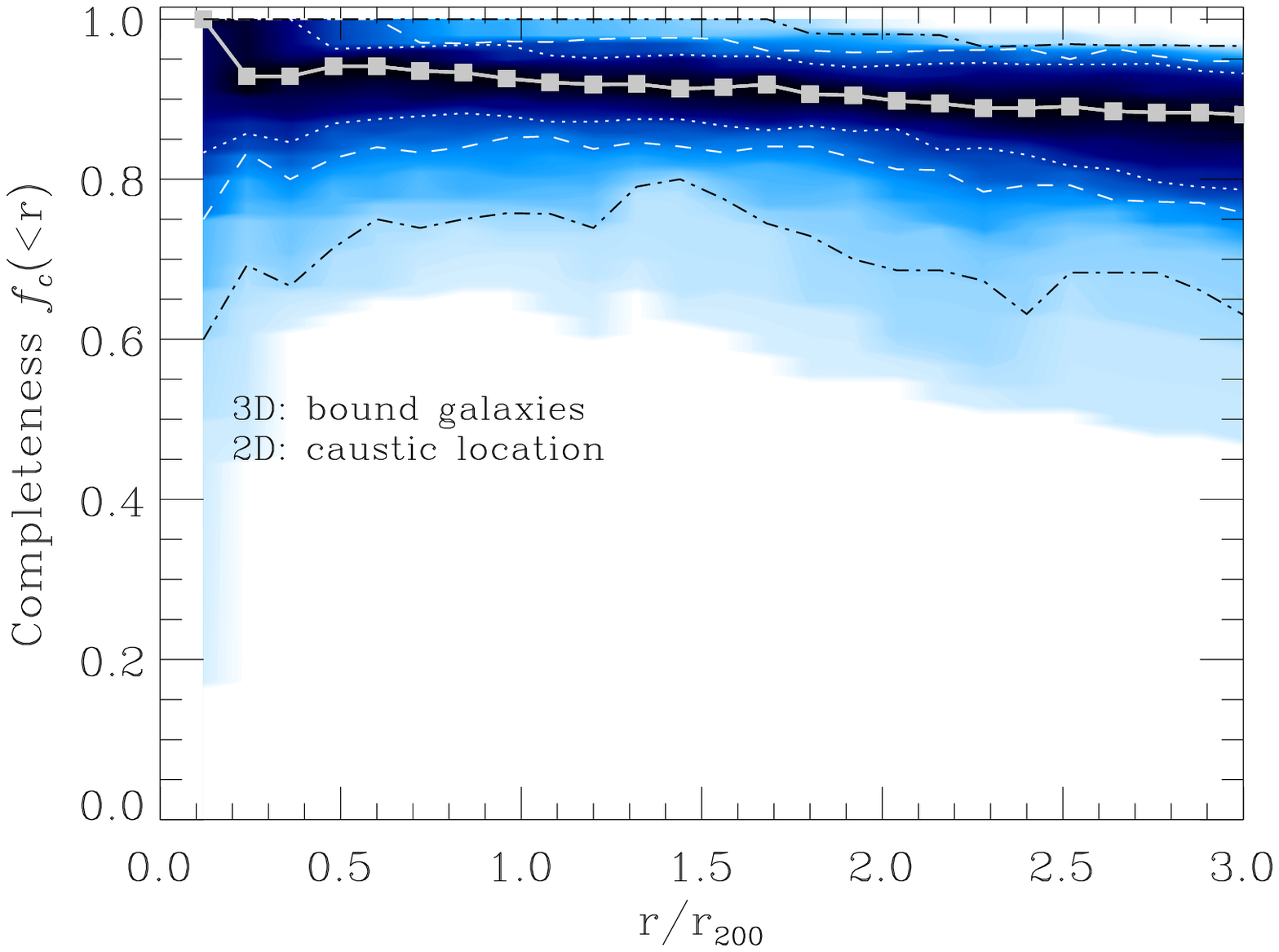}
\includegraphics[angle=0,scale=.55,trim=0 0 0 0]{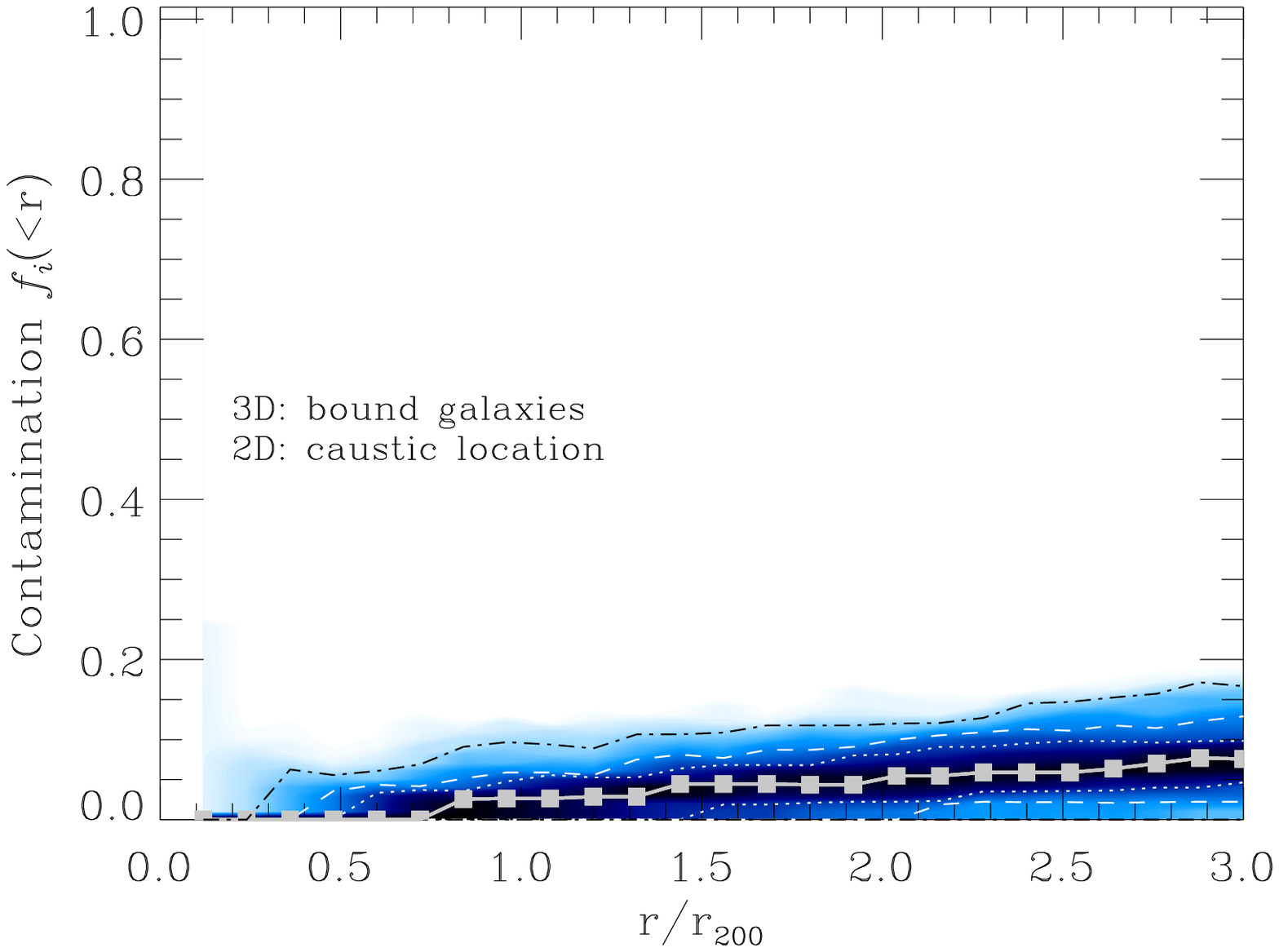}
\includegraphics[angle=0,scale=.55,trim=0 0 0 0]{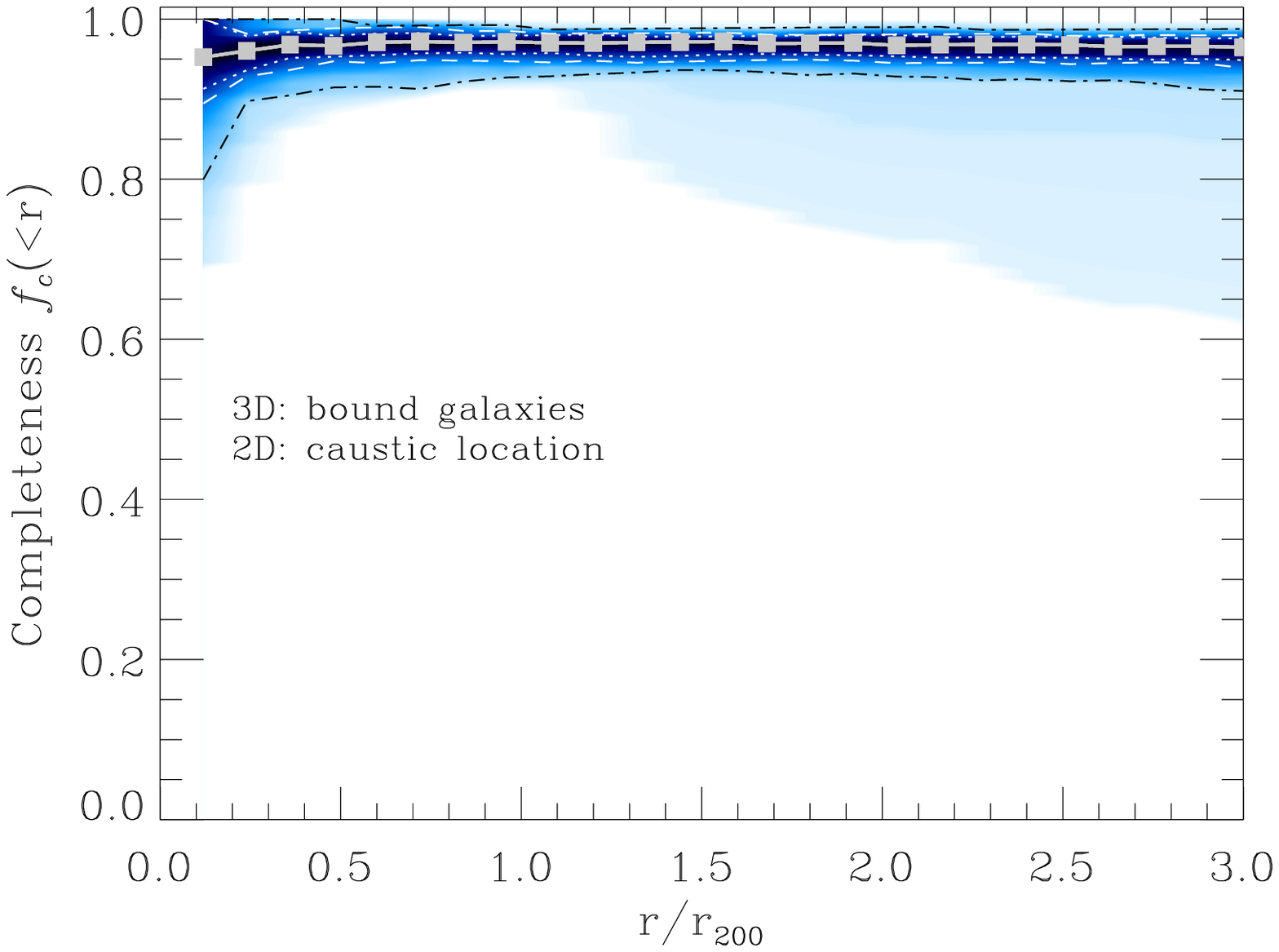}
\includegraphics[angle=0,scale=.55,trim=0 0 0 0]{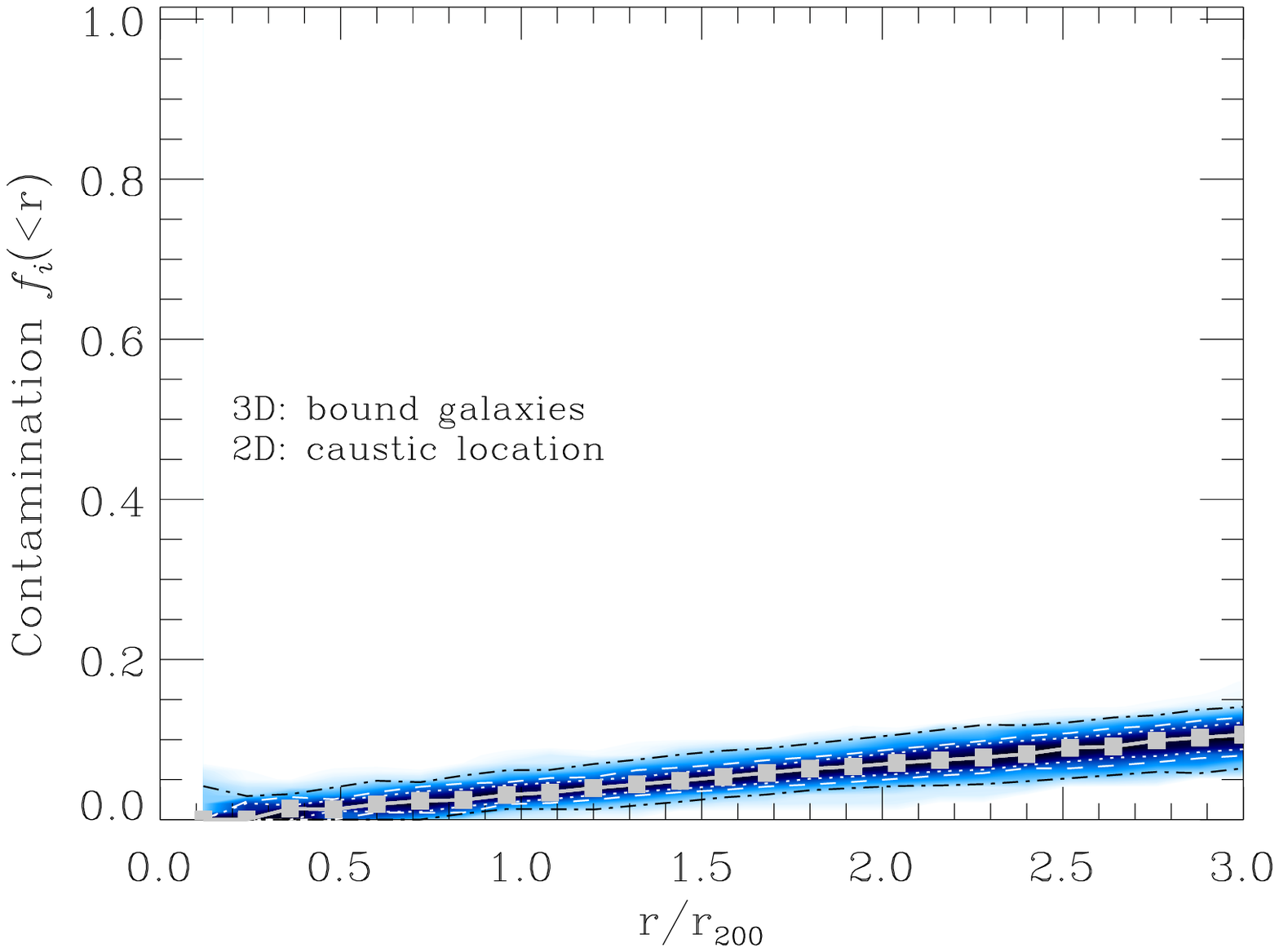}
\caption{Cumulative completeness $f_c$ (left panels) and contamination $f_i$ (right panels),
where the 3D members are the bound galaxies and the 2D members are the galaxies within the caustics, for the samples built from the stacked cluster with $N=50$ (upper panels) and $N=200$ (lower panels) particles 
within $3r_{200}$. The shaded areas and lines are the same as in Figure \ref{fig:evel-fg}.}
\label{fig:stacked100-bound-caus}
\end{figure*}

To quantify these effects, we stack our 3000 mock catalogs and randomly choose particles in the 
catalog until we obtain a given number $N$ of particles within 
$3r_{200}$ from the cluster center in real space. The stacking was done by scaling the coordinates with $r_{200}$ and the 3D velocity dispersion of each cluster. Here we show the results for two extreme cases with
$N=50$ and $N=200$. 
We compile 100 different catalogs for each value of $N$ and we apply the procedure to determine
the main group of the binary tree and to locate the caustics on the
corresponding redshift diagrams. The completeness and contamination profiles are shown in the upper panels of 
Figure \ref{fig:stacked100-bound-caus} for $N=50$ and in
the lower panels for $N=200$. We only show
the case where the bound galaxies are the 3D members, and the 
caustic location is used to select the 2D members.
The profiles show that the effect of increasing
$N$ from $50$ to $200$ is not significant on the median completeness and contamination profiles, but the associated
spreads drop by at least a factor of four for the completeness and a factor
of two for the contamination (Table \ref{table:comp1000}). 

\begin{table*}
\begin{center}

\begin{tabular}{ccccc|ccc||ccc|ccc}
&     &\multicolumn{12}{c}{\textbf{bound galaxies}}\\
&     &\multicolumn{6}{c}{\textbf{$N=50$}} &       \multicolumn{6}{c}{\textbf{$N=200$}}\\
\textbf{Caustic}&     &       &      &    &\\
\textbf{location} &    &   $f_c-\sigma$ & $f_c$  & $f_c+\sigma$     & $f_i-\sigma$ & $f_i$ & $f_i+\sigma$
      &  $f_c-\sigma$ & $f_c$ & $f_c+\sigma$  &   $f_i-\sigma$ & $f_i$ & $f_i+\sigma$\\
\cmidrule[0.05em](l ){3 -14}

&$r_{200}$  & 0.854  &  0.921  &  0.971   &  0.000  &   0.026  &  0.059
            & 0.946  &  0.970  &  0.982   &  0.021  &   0.034  &  0.052\\

&$2r_{200}$ & 0.812  &  0.898  &  0.959   &  0.000  &   0.055  &  0.100
            & 0.945  &  0.967  &  0.981   &  0.051  &   0.071  &  0.088\\

&$3r_{200}$ & 0.758  &  0.881  &  0.947   &  0.022  &   0.076  &  0.129
            & 0.939  &  0.965  &  0.980   &  0.080  &   0.106  &  0.128\\

\\
\end{tabular}

\end{center}
\caption{Cumulative completeness $f_c$ and contamination $f_i$ at $r_{200}$, 
$2r_{200}$, and $3r_{200}$ with their corresponding 1-$\sigma$ 
dispersion, with the bound particles as the 3D members and 
the particles within the caustics as the 2D members for the stacked cluster
with $N=50$ (left) and $N=200$ (right) particles within $3r_{200}$. }
\label{table:comp1000}

\end{table*}

\section{Mass Estimation} \label{sec:massestID}

In this section, we analyze the effect of our interloper removal 
methods on the estimation of the mass,
because interlopers have a non-negligible impact on the mass estimation,
especially at large radii, where interlopers can cause an overestimate
of the mass as large as a factor of three \citep{per90}. 

We consider the three standard methods described in \citet{HTB}: 
the virial, the average and the median mass estimators. All estimators assume
that the galaxies have equal mass and the system is in a steady state. The virial mass estimator is
\begin{equation}
M_{VT}= \frac{3\pi N}{2G} \frac{\sum_i v^2_{{\rm los} \,i}}{\sum_{i<j}1/R_{\perp,ij}}
\end{equation}
where $N$ is the number of galaxies with measured redshifts, $v_{\rm los}$ is the line-of-sight velocity and $R_{\perp,ij}$ is the projected separation between galaxy $i$ and
galaxy $j$. 
We do not add the surface pressure term in this analysis, because the correction that it introduces is
expected to be smaller than $10$\% \citep{Rin07}.

The median mass estimator is supposed to be less sensitive to interlopers, because
if interlopers in velocity populate the tails of the distribution of the quantity $(v_{{\rm los}\,i}-v_{{\rm los}\,j})^2 R_{\perp, ij}$, which is estimated for each of the $N(N-1)/2$ pairs, 
the median of this quantity is a more robust
estimate than the mean. The mass is thus 
\begin{equation}
M_{Me}= \frac{f_{Me}}{G}\, {\rm med}\{(v_{{\rm los}\,i}-v_{{\rm los}\,j})^2 R_{\perp, ij}\}\, ,
\end{equation}
where the coefficient $f_{Me}=6.5$ is calibrated with $N$-body simulations \citep{HTB}.

If we take the mean, rather than the median, we obtain the average mass estimator
\begin{equation}
M_{\rm AV}= \frac{3 f_{\rm AV}}{GN(N-1)} \sum_i \sum_{i<j}(v_{{\rm los}\,i}-v_{{\rm los}\,j})^2 R_{\perp, ij}\, ,
\end{equation}
where $f_{\rm AV}=2.8$ is again calibrated with $N$-body simulations \citep{HTB}. As mentioned
above, one expects that the virial and average mass estimators are more sensitive to interlopers
than the median mass estimator. 

We apply these three estimators to our samples after removing the interlopers 
with our two different procedures:
(1) the caustic location and (2) the binary tree.

\begin{figure*}
\center
\includegraphics[angle=0,scale=.5,trim=0 -20 -30 0]{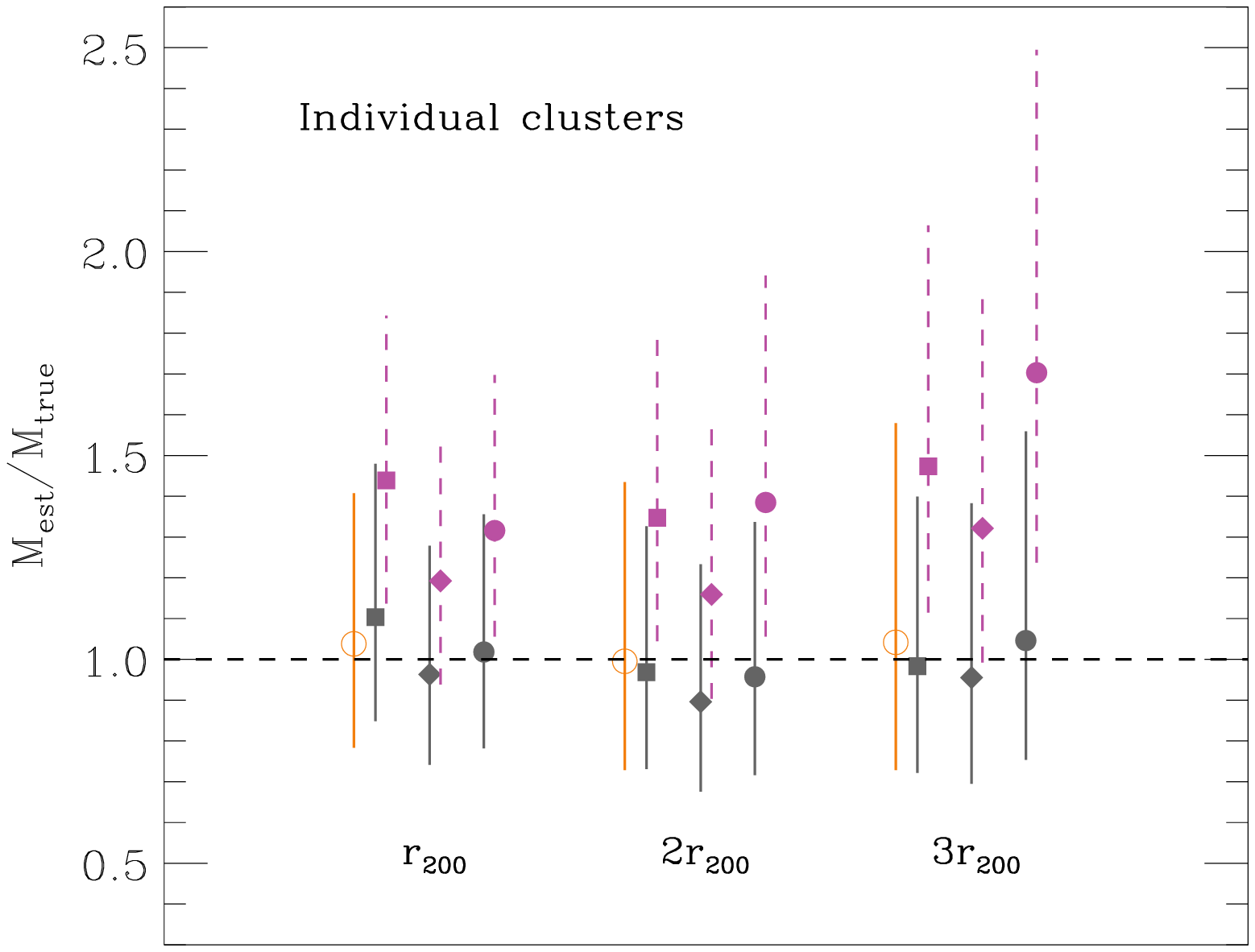} \\
\includegraphics[angle=0,scale=.5,trim=0 0 0 0]{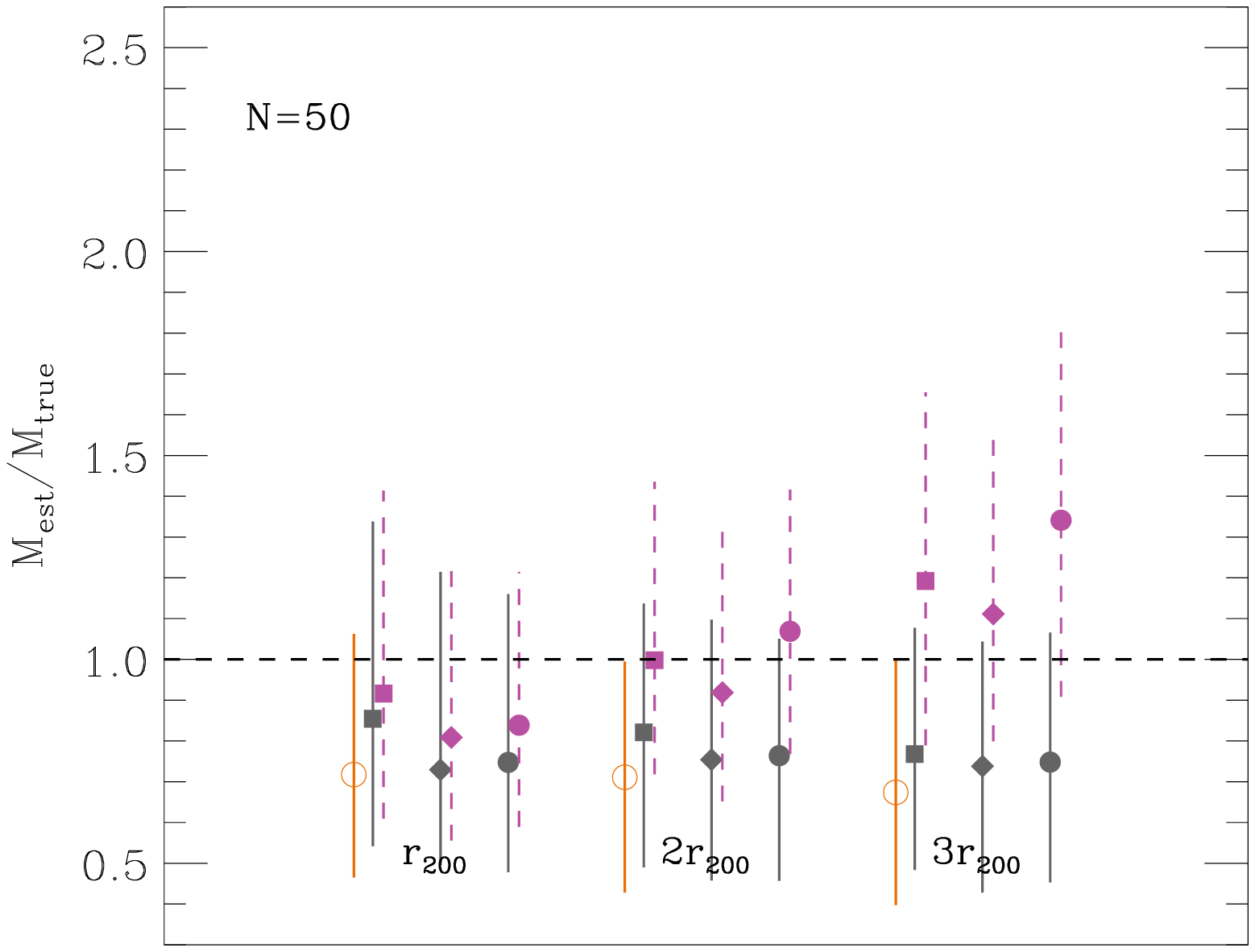}
\includegraphics[angle=0,scale=.5,trim=0 0 0 0]{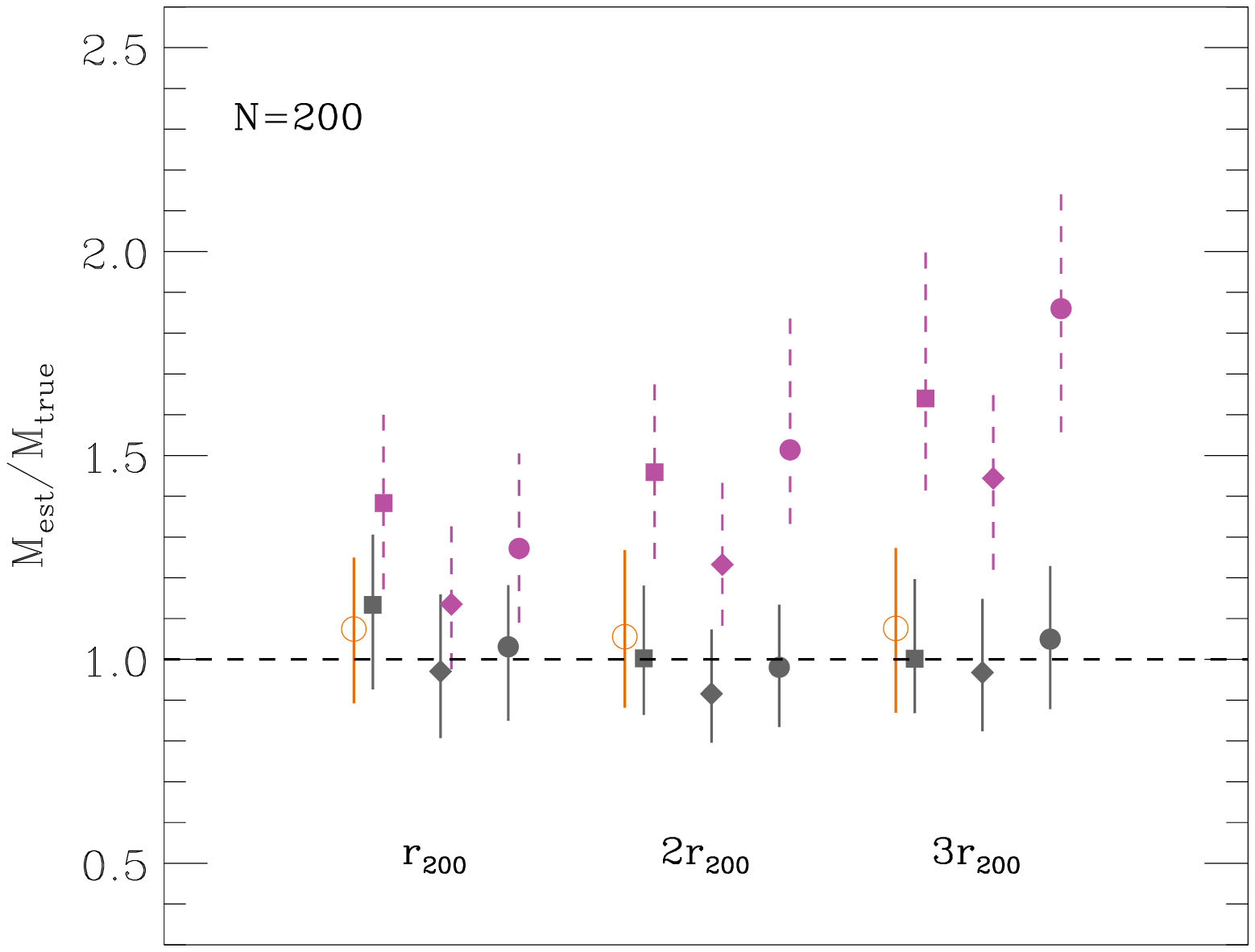}

\caption{Ratios $M_{\rm est}/M_{\rm true}$ between the cluster mass estimated with various
methods and the true mass derived from the $N$-body simulation. The mass estimators are applied
to individual clusters within $r_{200}$, $2r_{200}$ and $3r_{200}$ (top panel) and 
to the stacked clusters with $N=50$ and $N=200$ (bottom panels). 
$M_{\rm est}$ is estimated with the virial theorem (squares), the median (diamonds) or the average (solid circles) mass estimators; the interlopers are removed with the caustic location (gray), or the binary tree (violet). 
The error bars show the 68\% range of the distributions. The open circles show the mass estimated
with the caustic technique on the full particle sample, because the
caustic technique mass is not affected, in principle, by the presence of interlopers.}
\label{fig:massr200}
\end{figure*}

The top panel of Figure \ref{fig:massr200} shows the results of applying the mass estimators 
to individual clusters  within $r_{200}$, $2r_{200}$ and $3r_{200}$. On average, the mass estimate is unbiased when the member galaxies are identified with the caustic location, whereas it is
biased high by at least 20\% when the member galaxies are extracted from the main group of the binary tree. This result confirms our expectation that the caustic location removes interlopers
more efficiently than the binary tree procedure. 
As expected, the median mass estimator is the less sensitive method to the presence of
interlopers: in fact, it yields the values of $M_{\rm est}/M_{\rm true}$ closest
to one when the interlopers are removed with the less efficient binary tree procedure.

For comparison, we also show the mass estimated with the caustic technique applied to the
full sample of particles, because, in principle,
the technique is not affected by the presence of interlopers.
With all the estimators, 
the spread increases with radius. For the caustic technique, this increase
derives from the smaller number of galaxies available for locating the caustics. 
For the other estimators the system is required to be in virial equilibrium, that 
does not necessarily hold 
at radii larger than $r_{200}$; therefore, at these radii, the three standard estimators
are more likely to return an incorrect mass.

In addition, part of these spreads derives from the assumption of spherical symmetry. 
The bottom panels of Figure \ref{fig:massr200} show the mass estimates of the
stacked clusters with $N=50$ and $N=200$. The spreads for the case
$N=50$ are slightly smaller than in the case of individual clusters (upper panel of Figure \ref{fig:massr200}),
despite the fact that, on average, individual clusters have 185 galaxies within $3r_{200}$, a factor
of 3.7 larger than the $N=50$ stacked cluster (bottom left panel of Figure \ref{fig:massr200}). 
In the stacked clusters, 
the assumption of spherical symmetry is basically correct, therefore
the spread only derives from the sample size. In fact, the spreads further reduce by a factor 
of roughly $25$\% in the case of the $N=200$ stacked cluster (bottom right panel of Figure \ref{fig:massr200}). 

We finally note that in the case $N=50$, the mass estimate is biased low by 20\%. In
this case, in fact, the number of galaxies within $r_{200}$ is only $27$, on average, 
and the velocity field is too poorly sampled to return a correct mass. 

\section{Discussion}
\label{sec:disID}

Proper estimates of the mass of galaxy clusters and of the properties
of their galaxy population depend on the accurate separation between 
the cluster members and those galaxies that appear projected in the cluster field of view but
are not dynamically linked to the cluster.

Numerous methods to identify and remove interlopers in galaxy clusters
have been suggested in the literature. The algorithms are 
based either on the line-of-sight velocity separation of the galaxy from the cluster
center alone or on both the velocity and the projected separations. 
The former class of algorithms is suited for galaxy samples that
only survey the central regions of the clusters. 
These algorithms include the $3\sigma$ clipping method \citep{yah77},
which assumes that the velocity distribution is close to Gaussian,
the gap method  \citep{zab90,bee90}, and the adaptive kernel method \citep{pis93}. 
However, not all interlopers have large velocity separations 
from the cluster, as Figure \ref{fig:rediagbound} illustrates. These 
interlopers are difficult to identify and 
can generate a rather counterintuitive systematic error: they 
can cause a slight underestimate, rather than 
an overestimate, of the cluster velocity dispersion \citep{cen97, diaferio99, bivianoetal06}.

Rather than iterating over the velocity dispersion, like the $3\sigma$ clipping method does,
one can iterate over the virial mass: at each step, one removes the galaxy 
that causes the largest mass variation \citep{per90}.  
The projected and virial mass estimators \citep{HTB} are sensitive
to the presence of interlopers in different ways; comparing their
mass values can also be used to identify interlopers \citep{woj08}.
Iterating over the mass provides
more robust results than iterating over the velocity dispersion, because 
interlopers can 
affect more the estimate of the size of the cluster, which enters the mass estimate, 
than the velocity dispersion \citep{diaferio99}.
 
When galaxy catalogs survey large cluster regions, 
the methods described above can be extended and 
applied to galaxy subsamples separated into bins of projected  distances
to the cluster center \citep{fad96}. Thus, the velocity distribution 
assumed to be Gaussian at each radius can have different widths at different
radii \citep{pra03}, or the velocity dispersion in the $3\sigma$ clipping method
can be derived at different radii by solving the Jeans equation for a steady state system 
and isotropic galaxy orbits \citep{lok06}.

A step forward an interloper
rejection algorithm based on a dynamical approach derives from the following consideration: 
from an extensive galaxy sample 
we can actually extract information on the dependence of the escape velocity
on the clustrocentric distance \citep{har96}. 
Based on this idea, most algorithms
first estimate the mass profile by assuming dynamical equilibrium, and then, from these
mass profiles, derive the escape velocity as a function
of the projected distance to the cluster center.
The final solution thus must be obtained by iteratively removing the
identified interlopers until the mass profile converges.

Here, we have shown the performance
of the caustic technique, used as an interloper
rejection algorithm. The technique only uses the number density distribution of galaxies
in the redshift diagram to estimate directly the escape velocity profile from the system.
Unlike the methods mentioned above, the caustic technique relies neither on the
assumption of dynamical equilibrium nor on the estimate of the mass profile.
Therefore, the technique does not require any iteration; in addition, 
the estimate of the cluster
mass profile is a further step that is unnecessary for identifying the interlopers 
and it is a step that we have not taken here.

In the presence of extensive surveys, interloper rejection algorithms
based on the estimate of the mass or on the escape velocity usually perform better than
algorithms solely based on the velocity distribution \citep[e.g.,][]{woj07, woj08, whi10}.
\citet{woj08} perform an extensive comparison of a number of different 
algorithms. They conclude that the method by \citet{har96} is the most effective
at removing interlopers, producing samples with average contaminations $f_i$ in the range $0.02-0.04$
within one projected virial radius.  

These values are in perfect agreement with our median $f_i=0.02$ (Table \ref{table:comp}). 
However, there are two noticeable differences between our analysis and theirs:
the dynamical state of the clusters and the sample extension.
The caustic technique is independent of the dynamical state of the cluster, and, 
in fact, we only adopt the cluster mass as the
criterion to build our sample of $100$ simulated clusters.
On the contrary, in their sample of $10$ simulated clusters, \citet{woj08} pay particular attention to  
only include relaxed systems that have no sign of ongoing mergers, because
the \citet{har96} method requires dynamical equilibrium to be effective.
Merging clusters require more sophisticated approaches (see, e.g., \citealt{wegner11}  
and references therein), like the caustic technique.

In addition, the dynamical equilibrium assumption clearly limits the analysis to projected
radii smaller than the virial radius, where the equilibrium is
expected to hold, whereas the caustic technique
enables the identification of interlopers to much larger radii. We find 
that, with the caustic technique, the median contamination increases to $f_i=0.047$ and $0.080$ at $2r_{200}$
and $3r_{200}$, respectively, with a median completeness $f_c$ that remains larger
than $\sim 0.95$. No other methods that remove interlopers in these
regions are currently available. 

When we use the caustic technique to identify interlopers, the virial
mass estimator returns a mass overestimated by 10\% within $r_{200}$,
similarly to the results of \cite{bivianoetal06}, who removed interlopers
with a combination of the gap procedure \citep{gir93} and
the method of den Hartog and Katgert \citep{kat04,har96} from mock clusters
with more than 60 members. This bias is not present when we use the median and average
mass estimators. At radii larger than $r_{200}$, where only
the caustic technique can be used to remove interlopers, all mass estimates
are unbiased.

\section{Conclusion}
\label{sec:conID}

The caustic technique identifies the escape velocity profiles of galaxy clusters
to radii as large as $3r_{200}$; we can thus estimate the cluster
mass in regions where the cluster is not necessarily in dynamical 
equilibrium. The performance of the caustic method as a mass estimator
has been tested on both $N$-body simulations \citep{diaferio97,diaferio99,ser10} and real clusters 
\citep{diaf05b, geller2012}, regardless of their dynamical state: when we compare the caustic mass with the gravitational lensing mass in a combined sample of 22 clusters, the two estimates generally agree \citep{diaf05b, geller2012}.

Here, we have investigated an additional use of the caustic technique: the interloper rejection method. In this case, the technique relies only on the location of the caustics on the redshift diagram and makes no use of the mass profile of the cluster. We have tested the ability of the method to identify the cluster galaxy members by using $100$ galaxy clusters with mass $M_{200}\ge 10^{14}h^{-1} M_\odot$ extracted from a cosmological $N$-body simulation of a $\Lambda$CDM universe.  
Unlike the case of the mass estimate, where we compare the caustic technique with
gravitational lensing, we cannot test the interloper rejection method
on real clusters. However, the caustic technique is based on the hypothesis that clusters form by hierarchical clustering; wide observational evidence, based on X-ray and optical data, including gravitational lensing studies \citep[e.g.][]{diaferio08,bor11}, suggest that this hypothesis is well founded. Therefore we expect that $N$-body simulated clusters are a reasonable representation of real clusters and that
the results of our analysis can be safely applied to real clusters.

Our mock catalogs contain 1000 galaxies in the field of view of $12 h^{-1}$~Mpc on a side at the cluster location. The true 3D members, defined as the gravitationally bound galaxies, are compared to the galaxies identified as members with the caustic technique. We find a completeness of $f_c=0.95\pm 0.03$ within $3r_{200}$, whereas the contamination  
increases from $f_i=0.020^{+0.046}_{-0.015}$ at $r_{200}$ to $f_i=0.08^{+0.11}_{-0.05}$ at $3r_{200}$. The lack of spherical symmetry in clusters of galaxies causes most of the spread of the completeness and the contamination profiles. In fact, when applying the technique to samples built from a spherically symmetric stacked cluster the spreads decrease by at least a factor of two.
No other technique for the identification of the
members of a galaxy cluster provides such large completeness and 
small contamination at these large radii. 

The mass estimated with the virial theorem within $3r_{200}$, after removing interlopers in the case of individual clusters, is unbiased and is within 30\% of the actual mass. The use of the spherically symmetric stacked cluster decreases the spread to less than 10\%.

For the sake of clarity, we remind the systematic error that our interloper rejection method can introduce: the membership identification is based on identifying the caustic amplitude with the escape velocity {\it tout-court}, whereas the caustic amplitude, which we measure independently  of the knowledge of $g(\beta)$, of the mass profile and of the gravitational potential profile, actually is the escape velocity corrected by the factor $g^{-1/2}(\beta)$ (Equation \ref{eq:rig-pot}). The fact that we neglect this correction factor when we identify the caustic amplitude with the escape velocity can propagate in an incorrect separation of the cluster members from the interlopers. Our excellent results show that, despite this simplification, the caustic method can satisfactorily separate the members from the interlopers.

The increasing amount of data in clusters of galaxies \citep{geller11} 
requires adequate tool for extracting the information they contain
and properly comparing them with the output of the galaxy formation modeling that
is increasingly sophisticated \citep{sar12}. 

The caustic technique can provide accurate estimates of the dark matter distribution 
in the outer regions of galaxy clusters and information on the dynamical
connection between galaxies and clusters. The first piece of information
is relevant for our understanding of the formation of cosmic 
structure and can even constrain the properties of dark matter \citep{ser11} and 
the theory of gravity \citep{lam12}.

Determining the membership of galaxies in the outskirts of clusters is
unique to the caustic method. Applying the algorithm to a large
sample of clusters can provide the first accurate measure of 
how the gradients of properties of the cluster galaxy population, such as
color and star formation rate, merge into the field. In addition, it might
provide the first determination of galaxy membership in the filaments
surrounding clusters that represent the preferred path
of mass accretion \citep{pim04, col05, ara10,gon10}; 
this piece of information can thus enlighten the connection 
between the formation of galaxies and the large-scale structure.
In future work, we will investigate the reliability
of the caustic method in performing these measurements 
and assess the impact that these measures can have on the models of 
the formation of the cosmic structure.

\section*{ACKNOWLEDGMENTS}
We thank Giuseppe Murante, Stefano Borgani and the other members of our 
\citet{borgani04} collaboration for the use of our $N$-body simulations for this project.
We thank Margaret Geller for enlightening discussion and suggestions
and an anonymous referee whose comments prompted us to improve the presentation 
of the caustic technique and of our results. The simulations were carried out at the CINECA
supercomputing Centre in Bologna (Italy), with CPU time
assigned through a INAF-CINECA Key Project.
ALS acknowledges a fellowship by the PRIN INAF09
project ``Towards an Italian Network for Computational
Cosmology''.
Partial support from the INFN grant PD51 and the PRIN-MIUR-2008 grant \verb"2008NR3EBK_003"
``Matter-antimatter asymmetry,
dark matter and dark energy in the LHC era'' is also gratefully acknowledged.
This research has made use of NASA's Astrophysics Data System.

\bibliographystyle{apj}

\end {document}